\documentclass[a4paper,11pt]{article}

\usepackage{tikz}
\usetikzlibrary{backgrounds}
\usepackage{caption}

\usepackage{amsmath,amssymb}    %%ÃÂ»ÃÂ©ÃÃœÃÂ§Â·Ã»ÂºÃ
\usepackage[T1]{fontenc} % if needed
\usepackage[textsize=footnotesize,textwidth=2.5cm]{todonotes}
\newcommand{\wei}[2]{\textcolor{red}{#1}\todo[color=orange]{WS: #2}}

\usepackage{comment}

\usepackage{color}
\usepackage{graphicx}
\usepackage{subfigure}
\usepackage{cite}               %%ÃÃÃÃœÃÃÅŸÃ±ÃÅÂ±Ã€ÃÂ¯ÃÃ
\usepackage{hyperref}           %%Â³Â¬ÃÅœÅÃÂ±ÃÃÃ«

\numberwithin{equation}{section}   %%Â¹Â«ÃÅÂ°ÅœÅÃÂ±Ã ÂºÃ

\def \be {\begin{equation}}
\def \ee {\end{equation}}
\def \ba {\begin{array}}
\def \ea {\end{array}}
\def \bea{\begin{eqnarray}}
\def \eea{\end{eqnarray}}

\def \g {\gamma}

\def \G {\Gamma}
\def \d {\delta}

\def \m {\mu}

\def \p {\partial}
\def \f {\frac}

\def \nn {\nonumber}

\def \mc {\mathcal}

\def \mP{\mathcal P}

\def \hs {\hspace}

\def \inf {\infty}

\def \Tr {{\textrm{Tr}}}

\newcommand{\bra}{\left\langle}
\newcommand{\ket}{\right\rangle}

\begin{document}
\title{ R\'enyi Mutual Information in Holographic Warped CFTs}

\author{
Bin Chen$^{1,2,3}$,
Peng-Xiang Hao$^1$
and Wei Song$^4$
}
\date{}

\maketitle

\begin{center}
{\it
$^{1}$Department of Physics and State Key Laboratory of Nuclear Physics and Technology,\\Peking University, 5 Yiheyuan Rd, Beijing 100871, P.~R.~China\\
\vspace{2mm}
$^{2}$Collaborative Innovation Center of Quantum Matter, 5 Yiheyuan Rd, Beijing 100871, P.~R.~China\\
$^{3}$Center for High Energy Physics, Peking University, 5 Yiheyuan Rd, Beijing 100871, P.~R.~China\\
$^{4}$Yau Mathematical Science Center, Tsinghua University, Beijing 100084, P.~R.~China
}
\vspace{10mm}
\end{center}

\begin{abstract}

 % R\'enyi mutual information (RMI) has proven to be an important tool to probe the AdS/CFT correspondence beyond the classical order. In this article,
%RMI is further used to investigate the AdS/WCFT correspondence, an example of holographic dualities beyond AdS/CFT.  AdS/WCFT is a conjecture that Einstein gravity under the Compere-Song-Strominger boundary conditions is holographically dual to a warped conformal field theory (WCFT).
%We calculate R\'enyi mutual information in WCFT in the large central charge and large distance limit. By using the operator product expansion of twist operators up to level 3, we read the leading oder and the next-to-leading order RMI in the large central charge and small cross-ratio limits. The leading order result is furthermore confirmed using the conformal block expansion. Finally, we match the next-to-leading result to the 1-loop partition function in the bulk assuming XXXXX.

The study of R\'enyi mutual information (RMI) sheds light on the AdS/CFT corre-spondence beyond classical order. In this article, we study the R\'enyi mutual information between two intervals at large distance in two-dimensional holographic warped conformal field theory, which is conjectured to be dual to gravity on AdS$_3$ or warped AdS$_3$ spacetimes under Dirichlet-Neumann boundary conditions. By using the operator product expansion of twist operators up to level 3, we read the leading oder and the next-to-leading order RMI in the large central charge and small cross-ratio limits. %We compute the leading order RMI from the conformal block of the warped CFT and find agreement with the OPE method. We discuss the implications of RMI on the bulk semiclassical gravity.
The leading order result is furthermore confirmed using the conformal block expansion. Finally, we match the next-to-leading order result by a 1-loop calculation in the bulk.

\end{abstract}

\baselineskip 18pt
\thispagestyle{empty}

\newpage
\tableofcontents
\section{Introduction}

Holographic entanglement entropy opens a new window to study the AdS/CFT correspondence\cite{Maldacena:1997re}.
It has been proposed \cite{Ryu:2006bv,Ryu:2006ef} that the entanglement entropy of a subregion $A$ in the boundary CFT can be holographically computed by the area of the minimal surface $\Sigma_A$ in the bulk,
\be
S_A=\frac{{\rm Area}(\Sigma_A)}{4G_N}
\ee
where $\Sigma_A$ is homologous to the entangling surface$A$. This so-called Ryu-Takayanagi (RT) formula is reminiscent of the Bekenstein-Hawking entropy for the black hole. Actually, it has been shown in \cite{Lewkowycz:2013nqa} that the holographic entanglement entropy can be taken as a kind of generalized gravitational entropy and it can be computed by the classical action of the corresponding gravitational configuration. The quantum correction to the holographic entanglement entropy would then be related to the semi-classical gravitational effects in the bulk\cite{Barrella:2013wja,Faulkner:2013ana}.

The semiclassical gravity picture of  holographic entanglement entropy is most manifest in the AdS$_3$/CFT$_2$ correspondence. In AdS$_3$ gravity, under Brown-Henneaux boundary conditions,  the asymptotic symmetry group is generated by  two copies of the Virasoro algebra with central charge\cite{Brown:1986nw}
\be
c=\frac{3l}{2G_N},
\ee
where $l$ is the AdS radius. This suggests that a  consistent quantum gravity  on AdS$_3$ with a semiclassical limit should be holographic dual to a 2D CFT, and that such holographic CFTs should have a large central charge\cite{Strominger:1997eq}.  Further studies from modular invariance indicates that holographic CFTs also have a sparse spectrum of light states,  which further implies that  the vacuum is the dominant contribution to the torus partition at low temperature/energy limit \cite{Hartman:2014oaa}. In this setup, the holographic entanglement entropy  has been intensely studied. When the entangling surface in the CFT is just a single interval, the minimal surface in the bulk is just a geodesic, whose length matches with the single-interval entanglement entropy\cite{Ryu:2006bv}.  For the multi-interval case, the holographic entanglement entropy has been proved in \cite{Hartman:2013mia,Faulkner:2013yia} to be equal to the one in the CFT\footnote{For other aspects on the holographic entanglement entropy, see \cite{Ryu:2006ef,Headrick:2007km,Hubeny:2007xt} and the references in \cite{Rangamani:2016dms}.}.

Furthermore, other important entanglement quantities such as the R\'enyi entropy and the mutual information can be studied in the AdS$_3$/CFT$_2$ correspondence quantitatively as well. The R\'enyi entropy encodes rich information of the reduced density matrix and the entanglement. In general such quantities are  rather difficult to compute, as it requires the partition function of the field theory on a manifold of nontrivial topology and conical singularity. Holographically, one has to take into account the backreaction of a cosmic brane\cite{Lewkowycz:2013nqa,Dong:2016fnf}. However, in the semiclassical limit, the computation of the R\'enyi entropy is feasible  on both the field theory and the gravity sides, even for the multi-interval case. On the field theory side,   assuming that  the conformal block is dominated by the vacuum module at large $c$, one may use the recursive relation of the Virasoro block\cite{Hartman:2013mia} or the operator product expansion (OPE) of the twist operators\cite{Headrick:2010zt,Chen:2013kpa}
 to compute the correlation functions of the twist operators. In particular, the OPE of the twist operators allows us to read not only the leading-order(LO)  result in the $1/c$ expansion, which is linear in $c$ and corresponds to the semiclassical action of the gravitational configuration, but also the next-to-leading-order(NLO) result, which corresponds to the 1-loop correction to the holographic R\'enyi entropy\cite{Barrella:2013wja,Chen:2013kpa}, and even the next-next-to-leading-order (NNLO) result, which corresponds to the 2-loop quantum correction in gravity\cite{Chen:2013kpa,Headrick:2015gba}. On the gravity side, though the gravitational configurations corresponding to the higher genus Riemann surfaces resulting from the replica trick are hard to construct explicitly, their semiclassical action can be read via the Zograf-Takhtadzhyan action and the monodromy method\cite{Faulkner:2013yia}, and the 1-loop correction can be read by finding the Schottky uniformization\cite{Barrella:2013wja}.  The computations from the field theory and the bulk gravity agree remarkably well\cite{Chen:2013kpa,OPE,Chen:2014unl}. Furthermore,  R\'enyi entropy with higher $n$ also provides examples where the sparseness  condition does not necessarily imply that the identity block dominates the conformal block expansion, and new phase could appear \cite{Belin:2017nze, Dong:2018esp} \footnote{We thank A.Belin for pointing this out to us. }.

 Three-dimensional quantum gravity should be defined with respect to appropriate boundary conditions. Besides the usual Brown-Henneaux boundary conditions, there exist other sets of consistent asymptotic boundary conditions \cite{Compere:2013bya,Grumiller:2016pqb}. In particular, under the Comp\'ere-Song-Strominger(CSS) boundary conditions, the asymptotic symmetry group of AdS$_3$ gravity is generated by the Virasoro-Kac-Moody algebra\cite{Compere:2013bya}, which can be realized in a warped conformal field theory(WCFT)\cite{Hofman:2011zj,Detournay:2012pc}. This leads to the AdS$_3$/WCFT correspondence. In this correspondence, the holographic entanglement entropy presents some novel features due to the nontrivial boundary conditions. The single-interval entropy is not simply captured by the length of a geodesic with homologous condition in the bulk, but needs modification\cite{Castro:2015csg,Song:2016gtd}. It is definitely interesting to consider the R\'enyi entropy in the multi-interval case, which could shed new light on the AdS$_3$/WCFT correspondence.

In this note, we study the  R\'{e}nyi mutual information of two disjoint intervals in holographic warped CFT with the assumption that the vacuum module dominates the correlation function in the limit of large central charge.  We consider the case that the intervals are far apart so that the cross ratio $x$ is small. For warped CFTs, the vacuum module contains all the Virasoro and U(1) Kac-Moody descendants. We use the OPE of the twist operators to compute the partition function. We propose a warped conformal transformation from $n$-sheeted geometry to the plane such that the OPE coefficients can be read from the one-point functions of the quasi-primary operators. We consider the quasi-primary operators up to level $3$, and find the R\'enyi mutual information up to $x^3$, including both the leading-order(LO) result and the next-to-leading order(NLO) result in the $1/c$ expansion.

As a consistency check, we reproduce the leading-order result in the R\'enyi mutual information by studying the conformal blocks of the warped CFT in the large $c$ limit. The warped conformal block factorizes into a Virasoro block and a Kac-Moody block. Consequently, the leading order contribution of the R\'enyi mutual information can be put in the following form
\bea
 I^{(LO)}_n(x) &=& -\frac{(n-1)c}{12n}\log(1-x) \nn\\
 &&+(\mbox{holomorphic part of RMI in holographic CFTs}), \label{LOpic0}
 \eea
 where the first line comes purely from the Kac-Moody block, and the second line comes from the Virasoro block.

 The next-to-leading-order result is expected to match the 1-loop partition function of handlebody configurations, which can be read by summing over images from the 1-loop result of BTZ \cite{Castro:2017mfj}
 \be
Z_{CSS}^{1-loop}=\prod_{\gamma\in {\cal P}'}\prod_{l'=1}^\infty \frac{1}{1-q_\gamma^{l'}}\prod_{l=2}^\infty \frac{1}{1-q_\gamma^{l}}. \label{1loop0}
\ee
where $\mP'$ is a set of representatives of primitive conjugacy classes of the modified Schottky group $\G'$ which should now be compatible with the CSS boundary conditions.
Note that the CSS boundary conditions make a selection on both the linearized fluctuations and the allowed images we sum over. As was shown in \cite{Castro:2017mfj}, only the holomorphic vector modes and tensor modes contribute to the partition on a fixed background, consistent with the chiral Virasoro-Kac-Moody symmetry in the dual WCFT. In addition,  the saddle points should be obtained from images of given BTZ solutions generated by an $SL(2,\mathbb{R})\times U(1)$ quotient instead of $SL(2,\mathbb{R})\times SL(2,\mathbb{R})$.
Nevertheless, we provide an argument that the quotients acts the same way in the holomorphic sector, which enables us to perform the 1-loop calculation in the bulk, and find that it indeed agrees with the NLO result from WCFT calculation up to $x^3$. This is to be expected as WCFT is chiral, and the difference in the uniformization does not affect the holomorphic part.

The remaining parts of this paper are arranged as follows. In section 2, we set up the notations in warped CFT and discuss the operator product expansion. In section 3, we give the detailed prescriptions to calculate the mutual R\'{e}nyi information, involving the partition function and  twist operator expansion. We introduce the warped conformal transformation  relating the $n$-sheeted geometry in the single interval (on the plane) case to the flat warped geometry. In section 4, we study the large distance expansion of the R\'enyi mutual information in the holographic warped CFT. First, we classify the quasi-primary operators in the orbifold theory, give them explicitly up to level 3 and calculate their coefficients. Then we read the  R\'enyi mutual information and expand it in orders of $1/c$. In section 5, we provide a consistency check on the leading order result by computing the 4-point function of twist operators,  assuming that the conformal block is dominated by the vacuum Verma module in the large $c$ limit. This  allows us to reach the conclusion (\ref{LOpic0}). Furthermore we compare  the next-to-leading order result with the bulk computation (\ref{1loop0}) in section 6.  In section 7, we end with conclusion and discussions. Some technical details are collected in the appendices.

\section{The ABCs of WCFTs}

\subsection{Symmetries and spectral flow}
In this subsection, we provide a brief review of warped CFTs \cite{Detournay:2012pc,Song:2017czq}, starting with a definition on a Lorentzian cylinder, instead of the plane. The motivation comes from the holographic models where the bulk spacetime has a $SL(2,R)\times U(1)$ local isometry, and a $S^1\times R$ boundary topology \cite{Compere:2009zj, Compere:2013bya}.  Let us start with the so-called{ canonical cylinder} parametrized by $({\hat x},\, \hat y)$, with a canonical spatial circle $({\hat x},\,\hat y)\sim ({\hat x}+2\pi,\, \hat y)$.
Consider a two-dimensional local field theory on the $({\hat x},\hat{y})$ cylinder, with the following local symmetry.
\begin{equation}
{\hat x}\rightarrow f({\hat x}),\hs{3ex} \hat y\rightarrow \hat y+g({\hat x})\label{wtr}
\end{equation}
where $f({\hat x})$ and $g({\hat x})$ are periodic functions of $\hat x$. \eqref{wtr} is the defining property of warped CFTs, which are non-relativistic. The two directions are not on the equal footing in the sense that $\hat x$ allows reparametrization while $y$ does not.  The left moving stress tensor is denoted as $T(\hat x)$ and the left moving $U(1)$ current is denoted as $P(\hat x)$.
Under the warped transformation \eqref{wtr}, the stress tensor and the U(1) current transform as
\bea
P(f({\hat x}))&=&f({\hat x})'^{-1}(P({\hat x})+\frac{kg'({\hat x})}{2})\label{PTtrs}
\\
T(f({\hat x}))&=&f({\hat x})'^{-2}\Big(T({\hat x})-\frac{c}{12}s(f({\hat x}),{\hat x})-g'({\hat x})P({\hat x})-\frac{k g'({\hat x})^2}{4}\Big)
\eea
where $s(f,{\hat x})$ is the Schwarzian derivative,
\begin{equation}
s(f,{\hat x})=\frac{f'''}{f'}-\frac{3}{2}\left(\frac{f''}{f'}\right)^2,
\end{equation}
and the derivative is with respect to ${\hat x}$.
The Fourier modes %are defined as follows,
%\begin{equation}
%L_n=-\frac{1}{2\pi }\int \epsilon_n(z) T(z)dz,\hs{3ex}P_n=-\frac{1}{2\pi }\int \sigma_n(z)P(z)dz
%\end{equation}
%Choosing the test functions as \note{$\epsilon_n=e^{inx}$ and $\sigma_n=e^{inx}$ check conventions for the signs}, one gets
form the canonical warped algebra
\begin{equation}\label{ctr}
 [L_n,L_m]= (n-m)L_{n+m}+\frac{c}{12}n(n^2-1)\delta_{n+m,0}
\end{equation}
\begin{equation}
[L_n,P_m]= -mP_{m+n}
\end{equation}
\begin{equation}
 [P_n,P_m]= \frac{k}{2}n\delta_{n+m,0}
\end{equation}
We would like to classify the WCFTs by the central charge $c$, level $k$, as well as a vacuum charge $b$ on the{ canonical cylinder},
\be P_0^{vac} =i b\ee
which also determines the expectation value of the Viraosoro zero mode on the vacuum as
\be L_0^{vac}\equiv a-\frac{c}{24}, \quad a=-{b^2 \over k}.\ee
The relation between $L_0^{vac}$ and $P_0^{vac}$ is determined by the unitary bound as discussed in \cite{Detournay:2012pc}.

Now let us consider WCFTs on other geometries. According to the transformation rules \eqref{wtr}, under a shift
\be
{\hat x}\to {\hat x}, \quad \hat y\to \hat y-i\mu {\hat x},\,
\ee
the charges will transformation as
\begin{equation}
P_{n;\mu}=P_n+{ik\mu\over2}\delta_{n,0},\hs{3ex} L_{n;\mu}=L_n+i\mu P_n-({k\mu^2\over4}+\frac{c}{24})\delta_{n,0}
\end{equation} In particular, the vacuum charge will be shifted to $<P_0>=i(b+{k\mu\over2})$.
By choosing the parameter \be b=-{k\mu\over 2},\,\ee
we can therefore define a{ reference cylinder} with zero $U(1)$ charge, and a spatial circle $(\hat x,y')\sim{(\hat x+2\pi,\, y'-2\pi i\mu)}$.
We can make a smooth cover of the{ reference cylinder} by multiplying the circle by $n$ while keeping the direction invariant. Therefore the spatial circle becomes $({\hat x},y')\sim ({\hat x}+2\pi n,\,y'-2\pi in\mu).$

From either the{ canonical cylinder} or the{ reference cylinder}, we can define a ``plane'' by an exponential map in the Virasoro direction, while keeping the $U(1)$ direction unchanged.  To do so, we should view ${\hat x}$ as a null direction with ${\hat x}=t+\phi,\,\phi\sim \phi+2\pi$, and take the analytic continuation $t_E=it$. Since the spatial circle does not change the $y$ coordinate, $y$ is to be viewed as proportional to the time direction.
Namely, we define
the canonical plane $\mathcal{C}$ parameterized by $(z,\,{\hat y})$,
by the transformation
\begin{equation}
{\hat x}\to z=e^{i{\hat x}}=e^{-t_E+i\phi} ,\quad \hat y\to \hat y    \end{equation}
with points identified by \be  (z,\,y)\sim (z e^{2\pi i},\,\hat y),\ee
and non-vanishing vacuum charges \be  <P_0>_{\mathcal{C}}=ib,\quad  <L_0>_{\mathcal{C}}=a=-{b^2\over k}.\ee
The vacuum charges can be interpreted as inserting an operator $V$ at the origin, and the currents will be non-vanishing \be \langle T(z)\rangle_V={a\over z^2}, \quad \langle P(z)\rangle_V={b\over z}.\ee

Similarly,  the{ reference plane} $\mathcal{C}'$ parameterized by $(z,y')$,
is obtained from the the{ reference cylinder} by
\begin{equation}
{\hat x}\to z=e^{i {\hat x}}=e^{-t_E+i\phi},\quad y' \to y'.\end{equation}
 The{ reference plane} has the nice feature that all vacuum charges are zero, and therefore no operator insertion at the origin. The information of non-trivial vacuum charges is now encoded in the non-trivial boundary conditions
 \be  (z,\,\bar z)\sim (z \, e^{2\pi i},\,y'-2\pi i\mu ).\ee
The smooth cover of the{ reference plane} $\mathcal{C}'$ is made by  multiplying the circle by $n$ while keeping the direction invariant, then points will be identified as $(z,\, y')\sim (z e^{2\pi ni} ,\,  y'-2\pi in\mu )$. This property will help us understand the uniformazation map \eqref{unip}.
Note that neither the{ reference plane} nor the {canonical plane} are the complex plane in the usual sense, as they are not parameterized by a pair of holomorphic and anti-holomorphic coordinates which are complex conjugate to each other. Nevertheless, as a mathematical trick, we can still view $z$ as a holomorphic coordinate on the usual complex plane.

Furthermore, in this paper we will formulate physical questions on the physical cylinder with a thermal circle $(x,y)\sim (x+i\beta, y-i\bar\beta)$, or with a spatial circle $(x,y)\sim (x+L, y-\bar L)$. The physical cylinder can be mapped to the canonical cylinder with a rescaling and tilting, and furthermore mapped to the canonical plane or the reference plane.

%%%%%%%%%%%%%%%%%%%%%%%%%%%%%%
\subsection{Operator product expansion}
In this subsection we discussion the operator product expansion in WCFTs, and determine the coefficients in terms of the three-point coefficients.

Operators in WCFTs can be organized using primary operators and their Virasoro-Kac-Moody descendants.
Primary operators at the origin are labeled by the conformal weight $h$ and the $U(1)$ charge $Q$, with
\begin{equation}
L_n|O\rangle=0,\ \ P_n|O\rangle=0, \forall n>0\\
\end{equation}
and the descendants are linear combinations of $L_{-1}^{N_1}L_{-2}^{N_2}\cdots P_{-1}^{N_1}P_{-2}^{N_2}\cdots|O\rangle$.
Moving the operators from the origin by $L_{-1}$ and $P_{0}$, one gets the complete set of operator basis of the theory at any point.
Using the commutation relations \eqref{ctr}, all the Virasoro-Kac-Moody generators can be rewritten as the polynomials of $L_{-1},L_{-2},P_{-1}$.
Therefore it is
possible to further organize a warped conformal family by quasi-primary operators and their global descendants, which is the basis we use in this paper.
Using this basis,  the OPE ansatz can be written as
\begin{equation}
\phi(x_1,y_1)\phi^{\dagger}(x_2,y_2)=\sum_{k,n}c_k a_{k,n}\frac{e^{iQy_{12}}}{x_{12}^{2h-h_k-n}}\partial^n_{x_2}\phi_k(x_2,y_2).\label{ope}
\end{equation}
where $\phi_k$ are all the quasi-primary operators in the theory with zero $U(1)$ charge. The $x_{12}$ part is fixed by the conformal weight. The ansatz is chosen such that $a_{k,n}$ only depends on the weights while $c_k$ contains the dynamical information of the theory.
 The summation over $k,n$ includes all the quasi-primary operators and their descendants, which form  a complete basis.

In the following we will determine the coefficients $c_k$ and $a_{k,n}$ using two and three point functions.
Two point functions of quasi-primary operators in the $SL(2,R)\times U(1)$ invariant vacuum can be fixed by the global symmetry \cite{Song:2017czq}
\begin{equation}
\bra\phi(x_1,y_1)\phi^{\dagger}(x_2,y_2)\ket=d e^{iQy_{12}}\frac{1}{x_{12}^{2h}}
\end{equation}
where $Q$ is the U(1) charge of $\phi$, $\phi^{\dagger}$ has the opposite charge to $\phi$, and $d$ is a normalization factor.
Using the OPE \eqref{ope} and taking derivatives on the two-point function of the neutral operator $\phi_k$, we obtain the following three-point function
\begin{equation}
\label{thrp1}\bra \phi(x)\phi^{\dagger}(1) \phi_k(0) \ket_x=\sum_n c_k d_k a_{k,n}e^{iQy_{12}}\frac{(-1)^n(2h_k)_n}{(x-1)^{2h-h_k-n}}.
\end{equation}
where $(h_k)_n\equiv (h_k)(h_k+1)\cdots(h_k+n-1) $ is the raising Pochhammer symbol.

On the other hand, three-point function can also be determined by global symmetry and can be rewritten in a general form
\begin{equation}
\bra \phi(x_1,y_1)\phi^{\dagger}(x_2,y_2)\phi_k(x_3,y_3)\ket=c_{\phi\phi^\dagger\phi'}e^{iQy_{12}}\frac{1}{x_{12}^{2h-h_k}}\frac{1}{x_{31}^{h_k}}\frac{1}{x_{23}^{h_k}}\label{thrpf}
\end{equation}
where  $c_{\phi\phi^\dagger\phi_k}$ is the three-point coefficient, characterizing the dynamics of the theory.
\eqref{thrpf} can be expanded in terms of $x-1$,
\begin{equation}
\label{thrp2}\bra \phi(x)\phi^{\dagger}(1)\phi_k(0)\ket=c_{\phi\phi^\dagger\phi_k}\frac{e^{iQy_{12}}}{(x-1)^{2h-h_k}x^{h_k}}
=\sum_n c_{\phi\phi^\dagger\phi_k} e^{iQy_{12}}\frac{(-1)^n(h_k)_n/n!}{(x-1)^{2h-h_k-n}}.
\end{equation}
%\begin{equation}
%x^{-h_k}=(1+(x-1))^{-h_k}=(-1)^n(h_k)_n(x-1)^n/n!.
%\end{equation}
Comparing the two expressions \eqref{thrp1} and \eqref{thrp2}, we find
\begin{equation}
c_k=\frac{c_{\phi\phi^\dagger\phi_k}}{d_k},   \hs{3ex} a_{k,n}%=\frac{1}{n!}\frac{(-h_k)(-h_k-1)\cdots(-h_k-n+1)}{(-2h_k)(-2h_k-1)\cdots(-2h_k-n+1)}
=\frac{(h_k)_n}{n!(2h_k)_n}.\label{akn}.
\end{equation}

\subsection{Comments on holographic WCFTs}
Similar to their CFT cousins, WCFTs are specified by the operator spectrum labelled by $h_{{\phi_k}}, \,Q_{\phi_k}$ and OPE coefficients $C_{ijk}$.  It is interesting to ask what are the necessary conditions for a WCFT to be holographically dual to quantum theories of gravity with a semiclassical limit.
From the asymptotic symmetry analysis \cite{Compere:2013bya}, we learn that holographic WCFTs usually have: i) a large central charge, and ii) a negative level. In addition, in order that the entropy formula \cite{Detournay:2012pc} to be valid for large central charge and finite temperature, we expect iii) a sparse spectrum for the light operators, similar to \cite{Hartman:2014oaa}. However, a precise statement for the sparseness condition is not yet spelled out \footnote{A detailed analysis is in process in \cite{yu}.}. Similar to holographic CFTs, we also expect two closely related but non-equivalent conditions to the sparseness condition. One is that  iv) the vacuum block dominates the warped conformal block expansion, and the other is that v) the WCFT is maximally chaotic \cite{Apolo:2018oqv} \footnote{ See \cite{Perlmutter:2016pkf} for a discussion between the last three conditions for holographic CFTs. We thank L. Apolo for discussions on this point. }.
Throughout this paper, we explicitly assume condition  i) a large central charge and iv) vacuum dominance for holographic WCFTs.

\section{R\'enyi mutual information in WCFT }
Quantum entanglement plays a central role in many fields of physics. It is interesting to
use  measures of entanglement to probe WCFTs and their holographic duals.
Single interval entanglement entropy and R\'{ e}nyi entropy in WCFT was first calculated in \cite{Castro:2015csg} and revisited in \cite{Song:2016gtd} by generalizing the Rindler method \cite{Casini:2011kv} and the warped Cardy formula \cite{Detournay:2012pc}. A more general procedure of the generalized Rindler method was proposed in \cite{Jiang:2017ecm}.  In \cite{Song:2016gtd},  an additional parameter was introduced which is essential to match the bulk calculation in AdS$_3$ and warped AdS$_3$ spacetime \cite{Song:2016gtd} and in lower spin gravity \cite{Azeyanagi:2018har}.
Entanglement entropy on excited states has also been discussed in \cite{Apolo:2018oqv}.

While the aforementioned results are classical and for single intervals, in this paper we will go one step further by calculating the R\'{e}nyi mutual information for two disjoint intervals and obtain both the classical result and the leading quantum corrections.
In this section, we revisit the calculation of R\'enyi  entropy for single interval in WCFT and describe the strategy to calculate R\'enyi mutual information by adapting the OPE method developed in the context of CFT$_2$ \cite{Chen:2013kpa,Chen:2014unl}.

\subsection{Twist operator and uniformization }
In this subsection, we revisit the calculation of R\'enyi  entropy for single interval in WCFT and convert the previous Rindler transformation to a uniformization map, which facilitates later calculations of R\'enyi mutual information in Euclidean language.

Entanglement entropy measures the correlation between the subregion and its environment. Consider a spacelike subregion $A$, separated with its environment by an entangling surface, which is a co-dimension 2 submanifold. The entanglement entropy of $A$ is defined to be the von Neumann entropy of the reduced density matrix\cite{nielsen2010quantum,petz2008quantum},
\begin{equation}
S_A=-\Tr_A\rho_A\log\rho_A
\end{equation}
where the reduced density matrix $\rho_A$ is obtained by taking the partial trace of the density matrix of the whole system,
\begin{equation}
\rho_A=\Tr_{\bar{A}}\rho.
\end{equation}

In quantum field theories, the computation of the entanglement entropy is quite difficult due to the infinite degrees of freedom and the non-local operator $\log\rho_A$. The usual way to compute the entanglement entropy is to apply  the replica trick\cite{Holzhey:1994we} by making $n$ copies of the original theory and glue them together cyclicly along the interval $A$. An orbifold theory is obtained by modding out $Z_n$ of the tensor product of $n$ copies of the original theory. Denote  the original manifold by $\Sigma$,  its  $n$-th smooth cover $\Sigma_n$,  and define R\'{e}nyi entropy as
\be
S_n=\frac{1}{1-n}\log\frac{\mathcal{Z}_n}{\mathcal{Z}^n}
\ee
where  $\mathcal{Z}_n$ is the partition function on $\Sigma_n$, and $\mathcal{Z}$ is on $\Sigma$.
The entanglement entropy can then be read from the $n\to 1$ limit  after proper analytic continuation
\begin{equation}
S_A=\lim_{n\rightarrow 1}S_n.
\end{equation}
To proceed, one may introduce twist operators $\sigma$ and $\tilde{\sigma}$ in the vacuum of the orbifold theory. %, which is obtained by modding out the $Z_n$ symmetry of the tensor product of $n$ copies of the original theory.
The twist operators are operators of co-dimension two, nonlocal in the spacetime $d>2$. In two dimensions, the twist operators  $\sigma$ and $\tilde{\sigma}$ are local operators, located at the endpoints of the subregions. Correlation functions on $\Sigma_n$ can be rewritten in the orbifold theory with twist operator insertions,
\begin{equation}
\bra O_i(x)\ket_{\Sigma_n} =\frac{\bra O_i(x)\sigma\tilde{\sigma}\ket_1 }{\bra\sigma\tilde{\sigma}\ket_1 }
\end{equation}
\begin{equation}
\bra (O_iO_j)(x)\ket_{\Sigma_n}=\frac{\bra O_i(x)O_j(x)\sigma\tilde{\sigma}\ket_1}{\bra\sigma\tilde{\sigma}\ket_1}
\end{equation}
where $\bra ...\ket_1$ denotes the correlation with the insertion of the twist operators in the orbifold theory, and $x$ denotes collectively the coordinates on $\Sigma_n$.
For CFT$_2$, the quantum numbers of the twist operator can be found by using a uniformization map from $\Sigma_n$ to the plane $\mathcal{C}$, the transformation rule of the stress tensor, and the Ward identity.
Then the partition function can be computed by the correlation function of the twist operators\cite{Calabrese:2004eu,Calabrese:2005zw}.

 Whereas the aforementioned method using twist operators in CFT$_2$ is more conveniently formulated in Euclidean signature, previous computations of R\'{e}nyi entropy for single intervals for WCFT \cite{Castro:2015csg, Song:2016gtd} have been based on the Rindler method in Lorentzian signature.
 One goal of this subsection is to  translate the Rindler transformation to the uniformization map in the Euclidean signature, which facilitates the calculation of the R\'{e}nyi mutual information later.

In the calculation of R\'{e}nyi entropy and entanglement entropy for single intervals in WCFTs \cite{Castro:2015csg, Song:2016gtd},  the key step is to find the generalized Rindler transformation \cite{Jiang:2017ecm} which maps entanglement entropy to thermal entropy. Consider a subregion $A$ on a manifold $\Sigma$,  a generalized Rindler transformation is a symmetry transformation of the theory which maps the domain of causality $\mathcal{D}$ of the subregion $A$ to a Rindler spacetime $\tilde{\Sigma}$ characterized by a thermal identification.  The modular flow generator is the generator of the thermal identification on ${\tilde {\Sigma}}$ rewritten on $\Sigma$, and is required to annihilate the vacuum, leave $\mathcal{D}$ invariant, and map $\p \mathcal D$ to itself.
As the Rindler transformation is a symmetry transformation, the partition function on $\tilde{\Sigma}_n$ and the one on  ${\Sigma}_n$ are equivalent up to a unitary transformation, and the thermal entropy on $\tilde{\Sigma}_n$ is hence the R\'{e}nyi entropy for  the subregion $A$ on the vacuum.
In general, thermal entropy is also difficult to calculate directly. However, for theories with nice modular properties, such as WCFT, a Carly-like formula can be derived \cite{Detournay:2012pc}, and the partition function on $n$-copied Rindler space $\tilde{\Sigma}_n$,  the thermal entropy on $\tilde{\Sigma}_n$, and hence the  R\'{e}nyi entropy for $A$ can all be calculated. One can also read the quantum numbers of the twist operators.

More explicitly, consider an interval $A$ bounded by two end points $(0,0)$ and $(l,0)$ on  a cylinder $\Sigma$ parameterized by $(x,\,y)$, with a thermal circle $(x,y)\sim (x+i\beta, y-i\bar\beta)$. Results for the spatial circle can be obtained by replacing $i\beta, i\bar\beta $ by $L,\bar L$, respectively.
With a change of convention $\alpha_{there}=2\pi\mu_{here}$ and a shift of the interval, the Rindler transformation in a WCFT  \cite{Song:2016gtd} can be written as
\bea\label{rindler}
\tanh {\pi {\tilde x}\over \tilde \beta}&=&{\tanh {\pi (x-{l\over2})\over \beta}\over \tanh {\pi l\over 2\beta}},\\
{\tilde y}+({{\bar { \tilde \beta}}\over  \tilde \beta}-{2\pi \mu \over \tilde \beta }){\tilde x}&=&y+({\bar\beta\over \beta}-{2\pi \mu\over\beta})x.
\eea
%\bea
%\tanh {\pi {\tilde x}_A\over \tilde \beta}&=&{\tanh {\pi (x-{l\over2})\over \beta}\over \tanh {\pi l\over 2\beta}}\\
%{\tilde y}_A+({{\bar { \tilde \beta}}\over  \tilde \beta}-{2\pi \mu \over \tilde \beta }){\tilde x}_A&=&y
%\eea
We would like to consider a manifold $\Sigma$ with zero temperature, which can be  taken as the plane limit of \eqref{rindler} with $\beta\to \infty, \, p\equiv {\bar \beta-2\pi \mu\over \beta} $ fixed.
%In particular, when $p=0$ and $b=-{2\mu\over k}$, we find that the plane limit of $\Sigma$ is nothing but the reference plane we discussed in section 2.
Note that the Rindler transformation maps the domain of dependence $\mathcal{D}$ to the Rindler space ${\tilde \Sigma }$, with a thermal circle $({\tilde x}_{}, \, {\tilde y}_{})\sim ({\tilde x}_{} +i\tilde \beta , \, {\tilde y}_{}-i{\bar{\tilde\beta}})$. %We also have a modular flow generator $k_t={\tilde\beta}\p_{\tilde x}-{\bar {\tilde\beta}}\p_{\tilde y}=2\pi\mu \p_y +$.
Making $n$ replicas corresponds to  a thermal circle  $(\tilde x, \, \tilde y)\sim (\tilde x +in\tilde \beta , \, \tilde y-in{\bar{\tilde\beta}}).$ %One can perform the replica trick in Lorentzian signature analysis using Schwinger-Keldysh construction, similar to the discussions for CFT in Lorentzian signature \cite{Dong-}.
\footnote{In this paper we use the exponential map and Wick rotation to obtain the Euclidean version in a heuristic way.  A more proper analysis in Lorentzian signature is to use the Schwinger-Keldysh construction, similar to the discussions for CFT in Lorentzian signature
\cite{Dong:2016hjy}. As was discussed in \cite{Jiang:2017ecm}, the Rindler time is given by $\tau_{A}=\pi({{\tilde x}\over\tilde \beta}- {{\tilde y}\over{\bar {\tilde \beta}}})$. The Rindler space is only a wedge of flat spacetime, and different patches can be obtained  by defining a single-valued Rindler time $\tau=\tau_{A}+{m\pi\over2}i,\, m\in{\mathbb Z}$, which can furthermore be obtained from $\tilde x={\tilde x}_A+{i m\tilde\beta\over4}, \,\tilde{y}=\tilde{y}_A-{i m{\bar{\tilde\beta}}\over4} $. In particular, all points in the complement $\bar{A}$ will be mapped to a Rindler space with $m=2$. The thermal circle corresponds to gluing the patch with $m=4$ to the patch with $m=0$ along $A$.
Making $n$ replicas corresponds to interpolating the different copies from $0\to1\to 2\to\cdots\to m=4n\to0$. See also \cite{Wen:2018mev} for related work.
We leave all possible subtleties related to analytic continuations to future work.  }
%In \cite{Castro:2015csg} and \cite{Song:2016gtd}, the Rindler space is further approximated by a torus with infinitely large spatial circle, and the R\'{e}nyi entropy can be determined using modularity of the torus partition function.
% It has been observed that the R\'{e}nyi entropy is independent of the choice of $\tilde \beta,\,{\bar{\tilde \beta} } $. Therefore we can set ${\bar{\tilde\beta}}=0$ without losing generality.
By redefining \be z=e^{{2\pi \tilde x\over n \tilde \beta }},\,\quad \hat y=y'-{2b\over k}\log z =\tilde y+{{\bar { \tilde \beta}}\over  \tilde \beta}\tilde x,\ee  the replicated Rindler space will be mapped to the canonical plane or the reference plane, and we get a uniformization map\footnote{ The uniformization map \eqref{unip} with $p=0$ was also found in \cite{Apolo:2018oqv}, where entanglement entropy on excited states and quantum chaos have also been discussed. } in the Euclidean signature,
%\be
%z=(\frac{w}{w-l})^{\frac{1}{n}},\hs{3ex}{\tilde y}-n\mu\log z=y'-({2b\over k} +n\mu )\log z=y-pw.\label{unip}
%\ee
\be
z=(\frac{w}{w-l})^{\frac{1}{n}},\hs{3ex}{\hat y}-n\mu\log z=y'-({2b\over k} +n\mu )\log z=y-pw.\label{unip}
\ee
where we have replaced  $x$ by $w$ in the plane limit, and the phase assignment is chosen to map points on the $k-$th sheet to a patch on the plane with $arg z\in [0, {2\pi\over n})$.
The one-point function on the replicated geometry $\Sigma_n$ will then become
\begin{eqnarray*}
 \bra T(w)\ket_{\Sigma_n} &=&\frac{l^2}{w^2(w-l)^2}{h_n\over n}+p\frac{l}{w(w-l)}{iQ_n\over n}-\frac{kp^2}{4}, \\
\bra P(w)\ket_{\Sigma_n}&=&\frac{l}{w(w-l)}{iQ_n\over n}-\frac{kp}{2}.
\end{eqnarray*}
Here the  conformal dimension and the charge of the twist operator in the orbifold theory are given by the vacuum charges and spectral flow parameters
\be\label{hnqn} h_n=n(\frac{c(n^2-1)}{24n^2}+\frac{a}{n^2}-\frac{b}{n}\m-\frac{k}{4}\m^2),\hs{3ex}Q_n=-i(b+\frac{nk\mu}{2}) \ee
where $k$ is the Kac level,  $c$ is the central charge of the Virasoro algebra, $a, b$ are the nonvanishing vacuum charges on the canonical plane as discussed in section 2.
The above quantum numbers for the twist operator as well as the one-point functions indeed agree with those of \cite{Song:2017czq} in the plane limit with $\beta\to \infty, \, p\equiv \lim_{\beta\to\infty}{\bar \beta-2\pi \mu\over \beta} $ fixed.

For generality, we have kept three parameters in the transformation.  The vacuum charges $a$ or $b$ (determined by only one parameter) on the canonical cylinder/plane is a defining property of the WCFT theory, the twist parameter $\mu$ introduces a shift in $y$ when $w$ goes around a branching point on $\Sigma$, and is responsible for the short distance behavior in the orbifold theory. The parameter $p$ can be viewed as an additional spectral flow parameter on $\Sigma_n$ and provides the constant pieces of the one-point functions on $\Sigma_n$. From purely WCFT analysis, the freedom in choosing the $U(1)$ directions seems to lead to these three free parameters.  For example, \cite{Castro:2015csg} set $\mu=0$, \cite{Song:2016gtd} kept both $b$ and $\mu$ aribtrary. However, from holography, $b$ and $\mu$ do not correspond to independent quantities in the bulk calculation either in AdS$_3$ \cite{Song:2016gtd} or warped AdS black holes \cite{Azeyanagi:2018har}.  They have to be chosen appropriately. Hence one may wonder whether $\mu$ and $b$ are physically independent from a purely WCFT perspective.
 Indeed, if we further require that the twist operator has zero charges as $n\to1$, the parameter $\mu$ should be chosen such that the twist operator has zero charge and conformal dimension. This further requirement sets $b=-{k\mu\over2}$ as was noticed in \cite{Apolo:2018oqv}, and leads to two independent parameters. In particular, when $p=0$, the plane limit of $\Sigma$ is just the reference plane. In the following, we will still use all the parameters, while keep in mind that there is only one free parameter in $a,\, b,$ and $\mu$.

Let us further comment on the plane limit. In this paper we will use the approach of operator product  expansion to calculate the R\'enyi mutual information.
Since the operator product  expansion captures the short distance behavior, there is no difference in the calculation of the coefficients,  no matter whether the interval is on a plane or on the cylinder. We choose to deal with them on the plane for simplicity. We should also note that a further spectral flow parameterized by $p$ has no effect on the two-point functions except the exponential part. In the case of neutral operators, the entire two-point function will not change at all. The final result of  R\'enyi mutual information will be the same for arbitrary $p$, since the exponential factors cancel each other out, which is consistent with the picture that the causal development of the interval is just a strip independent of $y$ in WCFTs which are non-relativistic \cite{Castro:2015csg,Song:2016gtd}. In the following discussions, we will set $p=0$ unless otherwise specified.

To summarize, we point out that the uniformization map \eqref{unip} has the following properties,
\begin{itemize}
\item The uniformization map \eqref{unip} is a warped conformal transformation.
\item
\eqref{unip} can be obtained from the Rindler transformation \eqref{rindler} by a plane limit on the right hand side, and an exponential map from the left hand side.
The conformal weight and charge of the twist operator agree with that found in \cite{Song:2017czq} using the Rindler method.
\item When $n=1$, the two end points on $\Sigma$ are mapped to the origin and $\infty$ on the canonical plane respectively, with an additional shift in the $U(1)$ direction.
The twist operator has zero quantum numbers $h_1=Q_1=0,$ with the choice $b=-{2\mu\over k}$. When $p=0,$ the physical plane $\Sigma$ is the reference plane.
\item For $n\neq 1$, the map is multivalued. In particular, when $p=0,$ going around the circle around $(0,0)$ and $(l,0)$ on $\Sigma_n$ becomes $({w\over w-l}e^{2\pi in}, \, y\sim y-2\pi in \mu)$, which enlarges the circle on the reference plane $(z,\,y')\sim (z e^{2\pi i},\,y'-2\pi \mu)$ while keeping the same direction. This is more obvious from the corresponding cylinders. Points on the $k$-th sheet on $\Sigma_n$ is mapped to a slice on the plane with $arg z\in [0, {2\pi\over n})$.
\item The uniformation \eqref{unip} in the Viraosoro direction is exactly the same as that of the holomorphic part of a CFT$_2$.
\end{itemize}

\subsection{R\'{e}nyi mutual information}\label{sec:RMI}
In this subsection, we give  a prescription of calculating R\'{e}nyi mutual information for WCFTs using the OPE of twist operators for two disjoint subregions.

 R\'{e}nyi mutual information is another important notion of quantum entanglement. For two disjoint subregions $A$ and $B$, the
  R\'{e}nyi mutual information is defined as%with respect to two subregions by
\begin{equation}
I_n=S_n(A)+S_n(B)-S_n(A\cup B)
\end{equation}
which can also be expressed by the correlation function of the twist operators
\begin{equation}
\label{RMIdef}I_n=\frac{1}{n-1}\log\frac{\bra\sigma\tilde{\sigma}\sigma\tilde{\sigma} \ket}{\bra \sigma\tilde{\sigma}\ket \bra \sigma\tilde{\sigma}\ket}
\end{equation}
Note that the twist operators are inserted at proper locations to separate the different subregions. $\sigma\tilde{\sigma}$ is just short for
\begin{equation}
\bra \sigma\tilde{\sigma}\cdots\ket=\frac{\bra \sigma_n\tilde{\sigma}_n\cdots\ket}{\bra\sigma_1\tilde{\sigma}_1\cdots\ket^n}
\end{equation}
which means that the reduced density matrix should be normalized properly. By taking the $n\to 1$ limit, we have the mutual information
\be
S(A,B)=S(A)+S(B)-S(A\cup B),
\ee
which is always positive due to the subadditivity property of entanglement entropy. The mutual information characterizes the entanglement between two subregions such that even if the two subregions are far apart, the mutual information is non-vanishing due to quantum correlations\cite{Wolf}.

In practice, it is hard to compute the R\'enyi mutual information as the $n$-folded geometry can have not only  nontrivial topology, but also singularities.
Nevertheless, one may apply the operator product expansion of the twist operators to read the partition functions in the large distance limit. In two dimensional CFTs, the twist operators are local primary operators such that one can use the technology of the OPE of primary operators to read the mutual information between two disjoint intervals\cite{Headrick:2010zt,Calabrese:2010he,Chen:2013kpa}. In particular, for holographic CFTs, one may focus on the vacuum module such that the quasiprimary operators can be classified level by level and the R\'enyi mutual information can be read in powers of the cross ratio and $1/c$.
For the 2D warped CFTs, we can apply the same strategy to calculate the four point function for twist operators and the R\'enyi mutual information. Namely, we will first write down an OPE ansatz for the twist operators, find the dominating terms in holographic WCFTs, and calculate each contributing term. We outline the main steps below.

We consider the R\'{e}nyi mutual information of two disjoint intervals, with the end points of $A$ at $(x_1,y_1)$, $(x_2,y_2)$, and the end points of $B$ at  $(x_3,y_3)$, $(x_4,y_4)$.
 The R\'enyi mutual information  \eqref{RMIdef} can be rewritten in terms of correlation functions of the twist operators,
we have
\begin{equation}\label{Indef}
I_n(x)=\frac{1}{n-1}\log\frac{F_n(x)}{F_1(x)^n}
\end{equation}
where
\begin{equation}\label{fndef}
F_n(x)=\frac{\bra \sigma_n(\infty)\tilde{\sigma}_n(1)\sigma_n(x)\tilde{\sigma}_n(0)\ket }{\bra \sigma_n(\infty)\tilde{\sigma}_n(1)\ket \bra\sigma_n(x)\tilde{\sigma}_n(0)\ket}
\end{equation}
Note that the $y$-dependent pre-factors in the $F_n(x)$ cancel each other. Since the exchanged operators are all neutral, there will be no exponential factor in the two-point functions. As a consequence, the  R\'enyi mutual information will be $y$-independent. Also, it  depends only on the warped conformal invariant cross ratio $x$
\begin{equation}
x=\frac{x_{12}x_{34}}{x_{13}x_{24}}
\end{equation}
where $x_i$'s are the endpoints of the two intervals.

Twist operators in WCFTs will also be viewed as quasi-primary operators with conformal weight $h_n$ and charge $Q_n$ \eqref{hnqn}. Their OPE ansatz follows  \eqref{ope},
\begin{equation}
\sigma_n(x_1,y_1)\tilde{\sigma}_n(x_2,y_2)=\sum_{k,m}c_k a_{k,m}\frac{e^{iQy_{12}}}{x_{12}^{2h_n-h_k-m}} \partial^m_{x_2}\phi_k(x_2,y_2).\label{OPEholo}
\end{equation}
The $\phi_k$ runs over every quasi-primary operators in the orbifold theory, and carries vanishing U(1) charge. Expanding the twist operators is equivalent to insert a set of complete basis in the four-point function of the twist operators.

Now the task is to determine the OPE coefficients $c_k$ for quasi-primary operators.
From \eqref{akn}, we learn that the coefficient $c_k$ of $\phi_k$ is determined by the three-point coefficient with the twist operators $c_k={c_{\sigma{\tilde \sigma}\phi_k}\over d_k}$, which can furthermore be read from its one-point function in the $n$-folded geometry of a single interval using
\begin{equation}
\langle \phi_k(x_3)\rangle_{\Sigma_n}={\langle \sigma (x_1){\tilde \sigma}(x_2) \phi_k(x_3) \rangle \over\langle  \sigma (x_1){\tilde \sigma}(x_2)\rangle  }
\end{equation}
where $\Sigma_n$ denotes the $n$-folded geometry.
Then we get
\be
c_k=\frac{1}{d_k}x_{12}^{-h_k}\lim_{x_3\rightarrow\infty}x_3^{2h_k}<\phi_k(x_3)>_{\Sigma_n}\label{ck}
\ee
%\note{what is $C_I$?}
Finally one-point functions of the quasi-primary operators can be calculated using the transformation laws \eqref{PTtrs} and the uniformization map \eqref{unip} which maps the $n$-folded geometry to the canonical plane.

 %This picture is consistent with \cite{Castro:2015csg}, in the sense that,
%\begin{equation}
%\bra\sigma_1\tilde{\sigma}_1\ket\neq 1
%\end{equation}
%However, for the holographic warped CFT which is dual to AdS$_3$ gravity under the CSS boundary conditions, the twist operator $\sigma_1$  have vanishing weight and charge\cite{Song:2016gtd}.

Using the OPE ansatz \eqref{OPEholo} and the coefficients \eqref{akn}\eqref{ck}, the function $F_n(x)$ can be expanded in terms of the global conformal blocks,
\begin{eqnarray}
\nonumber  F_n(x) &=& \sum_{\{\phi_k\}}d_k c_k^2x^{h_k}\sum_{m_1,m_2}a_{k,m_1}a_{k,m_2}x^{\frac{m_1+m_2}{2}}\partial^{m_1}_{x_1}\partial^{m_2}_{x^2}(x_1-x_2)^{2h_k}|_{x_1=1,x_2=0}\\
  &=&\sum_{\{\phi_k\}}d_k c_k ^2x^{h_k}\  _2F_1(h_k,h_k,2h_k,x). \label{Fn}
\end{eqnarray}
where the $n$ dependence is hidden in $c_k={c_{\sigma_n\tilde{\sigma}_n\phi_k}\over d_k}$, and $_2F_1(h_k,h_k,2h_k,x)$ is hypergeometric function.
The summation over $\{\phi_k\}$ is on the quasi-primary operators in the propagating channels.

For holographic WCFTs, as discussed in the end of section 2.2, we assume that the four-point function of the twist operators is dominated by the vacuum module, similar to holographic CFTs \cite{Asplund:2014coa}. Hereafter the summation (\ref{OPEholo}) in the OPE of the twist operators  is only over the quasi-primary operators in the vacuum module, which can be constructed in a systematical way. Note that a difference from usual holographic CFTs is that now the vacuum module is generated by the Virasoro and Kac-Moody creation operators acting on the vacuum.  As $x$ is the conformal invariant cross ratio, the large distance expansion corresponds to the small cross ratio. From the power $x^{h_k}$, we see that the quasi-primary operators of low scaling dimensions give the leading order contributions. Consequently, we can work out the contributions of the quasi-primary operators to the R\'enyi mutual information level by level.

%More details of the calculation can be found in Appendix A.

We end this section with a brief summary of the OPE methods. To to calculate the  R\'{e}nyi mutual information, one needs to deal with the partition function on the $n$-sheeted geometry, which is a formidable task due to the non-trivial topology. One may turn to the correlation functions of the twist operators instead. Using the OPE of the twist operators and only considering the vacuum module, it is feasible to read the R\'enyi mutual information of two disjoint intervals. The essential point is that the OPE coefficients of the quasi-primaries in the vacuum module can be read by their one-point functions in the replicated geometry. By applying the warped conformal transformation (\ref{unip}), the coefficients can be determined analytically. Then the main tasks are finding the operators in the orbifold theory and calculating their one-point functions in the $n$-sheeted geometry.

\section{R\'enyi  mutual information from OPEs}

In this section, we use the OPE method outlined in section 3.2 to study the R\'enyi mutual information of two disjoint intervals in holographic warped CFT.
As the mutual information can be expanded in powers of the cross ratio, the leading contribution is captured by the quasiprimary operators of low scaling dimensions. In this work, we would like to find the contributions up to $x^3$. This requires us to find all quasi-primary operators in the orbifold theory up to level 3 and compute their OPE coefficients.

%As discussed in section 2.3,  holographic warped CFTs are assumed to have a large central charge and a sparse light spectrum, and the vacuum module dominate the four point function calculation.  Furthermore, we are interested in the case that the two intervals are far apart such that we can do large distance expansion as well.

 %In the large $c$ limit, we can focus on the vacuum module, which consists of the  operators containing the holomorphic stress tensor and U(1) current in the holographic warped CFT.

%and we need the OPE coefficients of the twist operators with the quasi-primary operators.
%We want to deal with a theory which may have a holographic duality. With large c in mind, we focus on the composite quasi-primary operators containing stress tensor and U(1) current. This part will give universal contributions in all such theories. We start with the single copy of the theory.
\subsection{Quasi-primary operators in the orbifold theory}

As the first step, we should find the construction of the quasi-primary operators in the vacuum module in the original warped CFT.
The states in the vacuum module of the warped CFT are created by the Virasoro generators $L_{-m}(m\geq 2)$ and Kac-Moody generators $P_{-n}(n\geq 1)$ acting on the $SL(2,R)\times U(1)$ invariant vacuum.
The generating function of the quasiprimary operators in the vacuum at level $N$ is
\begin{eqnarray*}
  Z(q) &=&\frac{1}{1-q}\prod_{m=2}^{\infty}\frac{1}{(1-q^m)^2} \\
  &=& 1+q+3q^2+5q^3+10q^4+16q^5+29q^6+45q^7+75q^8+\cdots \\
   &=& \sum_{N=0}^{\infty}P(N)q^N.
\end{eqnarray*}
Here $P(N)$ is the partition function, which gives the number of operators  at level $N$. The numbers are as follows,
\begin{table}[h]
\begin{tabular}{|c|c|c|c|c|c|c|c|c|c|}
\hline
Level &  0 &  1 &  2 &  3 & 4&5&6&7&8 \\
\hline
 Number of operators &  1 &  1 &  3 &  5 &  10 &16&29&45&75 \\
\hline
 Number of quasi-primaries &  1 &  1 &  2 &  2 &  5&6&13&16&30 \\
\hline
\end{tabular}
\end{table}
\\We list the quasi-primary operators explicitly up to level 3.
\begin{itemize}
  \item Level 0, there is the identity operator $I$, corresponding to the vacuum state.
  \item Level 1, there is one quasi-primary operator, $-iP$, which is hermitian. It corresponds to the state $P_{-1}|0\rangle$.
\item Level 2, there are two quasi-primary operators. One is the stress tensor $T$, which corresponds to the state $L_{-2}|0\rangle$. The other one is $A=T+\frac{c}{k}(PP) $, which corresponds to the state $ (L_{-2} -\frac{c}{k}P_{-1}P_{-1})|0\rangle$.
    \item Level 3, there are two quasi-primary operators. One is $i(P^3)$, which corresponds to the state $P_{-1}P_{-1}P_{-1}|0\rangle$. The other one is $B=iP^3+ik(PT)$, which corresponds to the state $ (P_{-1}^3-kP_{-1}L_{-2})|0\rangle$.
\end{itemize}
We denote the normal ordered product of the operators $A$ and $B$ by $(AB)(w)$. The notation of normal ordered product we use is,
\begin{equation}
(AB)(w)=\frac{1}{2\pi i}\oint_w\frac{dz}{z-w}A(z)B(w),
\end{equation}
and
\begin{equation}
(P^3)=(PP^2)=(P^2P).
\end{equation}
Note that the OPE of $T$ and $P$ has singular terms, so that we cannot separate the contribution from the ones involving stress tensor and the ones involving U(1) current. We use the Schmidt orthogonalization process to make the quasi-primary operators at each level independent of each other  in the following calculation. One should also note that proper number of $\sqrt{-1}$'s should be included to make sure every quasi-primary operators above are hermitian.

%\subsection{n-Copy theory}
Let us turn to the construction of the quasi-primary operators in the vacuum module in the orbifold CFT.  Usually, the spectrum of a CFT includes the normal sector and the twist sector. Here we take a different point of view. Instead of classifying the operators according to their monodromy property, we classify the operators in the orbifold theory by the  operators in $n$-replicated theory, taking into account of the replica symmetry\cite{Calabrese:2004eu,Chen:2013kpa}.  The generating function of $n$-copied theory reads
\begin{eqnarray*}
  Z(q) &=&(\frac{1}{1-q}\prod_{m=2}^{\infty}\frac{1}{(1-q^m)^2})^{n} \\
  &=& 1+nq+\frac{5n+n^2}{2}q^2+\frac{14n+15n^2+n^3}{6}q^3+\cdots \\
   &=& \sum_{N=0}^{\infty}P(N)q^N,
\end{eqnarray*}
where the partition functions $P(N)$ give the number of operators at the level $N$. We get the number of quasi-primary operators by subtraction. The numbers are listed in the table below.
\begin{table}[h]
\begin{tabular}{|c|c|c|c|c|c|}
\hline
Level &  0 &  1 &  2 &  3 & 4 \\
\hline
 Number of quasi-primaries &  1 &  $n$ &  $\frac{n(n+3)}{2}$ &  $\frac{n^3+12n^2-n}{6}$ &  $\frac{n^4+26n^2+71n^2+22n}{24}$ \\
\hline
\end{tabular}
\end{table}

The explicit forms of the quasi-primary operators at each level are listed as follows.
\begin{itemize}
  \item Level 0, there is the identity operator $I$.
  \item Level 1, there is one kind of quasi-primary operators $-iP_i$. The subscript index $i$ labels the $i$-th copy of the theory  the operators belong to. There are $n$ such operators in total.
  \item Level 2, there are three kinds of quasi-primary operators. The first kind is $T_i$, and its  number is $n$. The second type is $A_i$, and its total number is also $n$. The last type is $P_iP_j$. We require $0\leq i<j\leq n-1$ to avoid double counting. The total number of such operators is $\frac{n(n-1)}{2}$. In sum, the total number of the operators at this level is $\frac{n(n+3)}{2}$.
  \item Level 3, there are six kinds of quasi-primary operators. They are $B_i$, $iP^3_i$, $-iT_iP_j$, $-iP_iA_j$, $iP_iP_jP_k$, and $-i(P_i\partial P_j-P_j\partial P_i)$ respectively. When the same kind of operators appear in the composite operators, their indices should be arranged increasingly in order to avoid repeatedly counting. The corresponding degeneracies of these operators are $n$, $n$, $n(n-1)$, $n(n-1)$, $\frac{n(n-1)(n-2)}{6}$, and $\frac{n(n-1)}{2}$ respectively. And the total number of the operators at this level is $\frac{n^3+12n^2-n}{6}$.
\end{itemize}
It is worth noting that not all of the quasi-primary operators can be written as the tensor product of the ones in the single copy theory. Such operators actually  belong to the twist sector of the orbifold CFT. %We also count the total number of quasi-primary operators we construct  to compare with the one calculated by generating function.

\subsection{Coefficients in the OPE}

To compute the function $F_n(x)$, we have to determine the OPE coefficients $c_k$ and the normalization factor $d_k$ in (\ref{Fn}). The normalization factors can be determined simply by algebraic relation, while the OPE coefficients can be determined by the
 the one-point functions of the above quasi-primary operators in the $n$-sheeted geometry with a single interval.  As now the quasi-primary operators are constructed by the stress tensor and the operator $P$, their  one-point functions are determined purely by the warped conformal transformation (\ref{unip}), similar to the case in usual large $c$ CFT\cite{Chen:2013kpa}.
The twist operator here is actually a composite operator including the standard one and the operator which generate the non-trivial vacuum. Since we has included such effect in the OPE coefficients, now we calculate the correlators of twist operators in the reference plane where the vacuum is trivial.
 We collect the main results of the OPE coefficients in this subsection, leaving some calculation details to the appendix \ref{OPEcoeff}. %One important subtle thing  is whether the one point functions are single-valued or not.
\begin{itemize}
  \item Level 0, the coefficients before the identity operator is 1 according to our normalization.
  \item Level 1, the quasi-primary operators are $-iP_i$, with
  \begin{equation}
  c_{-iP}=-i\frac{2b}{nk}-i\mu.
  \end{equation}
\item Level 2, for  $T_i$,
\begin{equation}
c_T=\frac{n^2-1}{12n^2}+\frac{2a}{cn^2}-\frac{2b\m}{cn}-\frac{k\mu^2}{2c}.\label{ct}
\end{equation}
For  $A_i$,
\begin{equation}
c_A=\frac{(2 b+k n \m)^2}{2 c k n^2}.
\end{equation}
For $P_iP_j$,
\begin{equation}
c_{P_iP_j}=(\frac{4b^2}{k^2n^2}+\frac{1}{2kn^2s_{ij}^2}+\frac{4b\m}{kn}+\m^2).
\end{equation}
\item Level 3, for $iP^3_i$,
\begin{equation}
c_{iP^3}=i\frac{(2 b+k n \m) \left(2 (2 b+k n \m)^2-k
   n^2+k\right)}{12 k^3 n^3}.
\end{equation}
For $B_i=iP^3_i+ik(PT)_i$,
\begin{equation}
c_{B}=\frac{i \left(n^2-1\right) (2 b+k n
   \m)}{12 k^2 n^3}.
\end{equation}
For $-iT_iP_j$,
\begin{equation}
c_{-iT_iP_j}=-i\frac{(2 b+k n \m) \left(-24 a+6 n \m
   (4 b+k n \m)-c n^2+c+6 s_{ij}^{-2}\right)}{12 c k
   n^3}.
\end{equation}
For $-iP_iA_j$,
\begin{equation}
c_{-iP_iA_j}=-i\frac{(2 b+k n\mu) \left((2 b+k n\mu)^2+k s_{ij}^{-2}\right)}{2 c k^2 n^3}.
\end{equation}
For $iP_iP_jP_k$,
\begin{eqnarray}
\nonumber c_{iP_iP_jP_k}&=&i\frac{ (2 b+k n\mu)^3
}{ k^3 n^3} + i\frac{ (k n\mu+2b)( s_{ij}^{-2}+s_{ik}^{-2}+
  s_{jk}^{-2})}{2 k^2 n^3}.
\end{eqnarray}
For $-i(P_i\partial P_j-P_j\partial P_i)$,
\begin{equation}
c_{-i(P_i\partial P_j-P_j\partial P_i)}=\frac{c_{ij}}{4kn^3s_{ij}^3},
\end{equation}
where
\begin{equation}
c_{ij}\equiv \cos\frac{\theta_i-\theta_j}{2n}=\cos\frac{\pi(i-j)}{n},\hs{3ex}s_{ij}\equiv\sin\frac{\pi(i-j)}{n}.
\end{equation}
\end{itemize}

%\subsection{Results}
With the coefficients above and the twist OPE ansatz, we can read the R\'enyi mutual information. It is worth noting that there is a cancellation between the numerator and denominator.

\subsection{R\'enyi entropy on the cylinder}
As a consistent check, we first consider the R\'enyi entropy of single interval with end points $(x_1,\,y_1),\,(x_2,\,y_2)$ on a cylinder $\Sigma$ with a spatial circle  $(x,y)\sim (x+L,y-\bar L)$. %To simplify the notations, we will use the notation $y_{12}=y_1-y_2, \,x_=x_1-x_2$ in this section.
The R\'{e}nyi entropy obtained in \cite{Castro:2015csg} and \cite{Song:2016gtd} can be put in the same form as  \begin{equation}
S_n=\frac{i}{1-n}(Q_n-nQ_1) (y_{12}-{\bar L\over L} x_{12})+\frac{1}{1-n}(nh_1-h_n)\log(\frac{L}{\pi\epsilon}\sin\frac{\pi x_{12}}{L})\label{Sncylinder}
\end{equation}
where  conformal dimension and the charge of the twist operator in the orbifold theory are given by the vacuum charges and spectral flow parameters
\be h_n=n(\frac{c(n^2-1)}{24n^2}+\frac{a}{n^2}-\frac{b}{n}\m-\frac{k}{4}\m^2),\hs{3ex}Q_n=-i(b+\frac{nk\mu}{2}). \ee
Note that the form of \eqref{Sncylinder} depends on the spectral flow parameters only through the conformal weight and charge of the spectral flow parameters.
Further requiring $h_1=Q_1=0$,  $\mu$ and $b$ are related by  \cite{Apolo:2018oqv},
\be b=-{k\mu\over 2}\label{mub}\ee
Then the quantum numbers of the twist operator become
\begin{equation}\label{dinv}
h_n^{inv}\equiv h_n-{Q_n^2\over nk} =\frac{c(n^2-1)}{24n},\hs{3ex}Q_n=(n-1)\frac{ik\mu}{2}
\end{equation}

%\note{ \begin{equation}
%S_n=\frac{i}{1-n}(Q_n-nQ_1) (y_{12}-{\bar L\over L} x_{12} -{Q_n^2\over nk}   \log(\frac{L}{\pi\epsilon}\sin\frac{\pi x_{12}}{L})  )+\frac{1}{1-n}(nh_1^{inv}-h_n^{inv})\log(\frac{L}{\pi\epsilon}\sin\frac{\pi x_{12}}{L})\label{Sncylinder}
%\end{equation}
%}
Now we show that the result  (\ref{Sncylinder}) can be reproduced using the OPE method outlined in section 3. For simplicity, let us first consider $\bar L=0$.
In terms of the twist operators, the R\'enyi entropy is read from
\begin{equation}
S_n=\frac{1}{1-n}\log\frac{\bra \sigma_n(x)\tilde{\sigma}_n(0)\ket}{\bra \sigma_1(x)\tilde{\sigma}_1(0)\ket^n}.
\end{equation}
After substituting the twist OPE ansatz into this formula and separating the contributions from the weight and the charge, it becomes
\begin{equation}
S_n=\frac{i}{1-n}(Q_n-nQ_1) y_{12}+\frac{1}{1-n}(nh_1-h_n)\log\frac{x_{12}}{\epsilon}+\frac{1}{1-n}\log\frac{G(n)}{G(1)^n}\label{Sncylinder2}
\end{equation}
where
\begin{equation}
G(n)=\sum_{\{\phi_k\}}c_kx^{h_k}\sum_{m}x^m a_{k,m}\partial^m_{x'}<\phi_k(x')>|_{x'=0}.\label{gn}
\end{equation}
comes from nonvanishing one-point functions on the cylinder. We use the following transformation to map the plane to the cylinder, %considering the vanishing spectrum flow parameter for simplicity.
\begin{equation}
z=e^{\frac{2\pi i}{L}x},\hs{3ex}{\hat y}=y+\mu \log z
\end{equation}
where $(z,{\hat y})$ are coordinates on the canonical plane, $(x,y)$ are coordinates on the cylinder $\Sigma$. As was argued in section 3.2, the parameter $\mu$ should be chosen as \eqref{mub} to make the twist operator chargeless at $n=1$. The expectation values of the quasi-primary operators up to level $2$ on the cylinder are respectively
\begin{equation}
\bra P(x)\ket=0, \hs{3ex}\bra T(x) \ket=\frac{\pi^2c}{6L^2},\hs{3ex}\bra P^2(x)\ket
  =\frac{\pi^2k}{6L^2},
\end{equation}
\begin{equation}
\bra A(x)\ket=0, \hs{5ex}\bra P_iP_j(x)\ket=0.
\end{equation} Note that all the above one-point functions on $\Sigma$ are independent of the vacuum charge $b$, provided that the parameter $\mu$ is chosen as \eqref{mub}.
After doing the summation in \eqref{gn}, we have
\begin{equation}G(n)=1+\frac{\pi^2 c}{6L^2}c_T n x_{12}^2+O(x^3).
\end{equation}
with \begin{equation}
c_T=\frac{2}{nc}h_n
\end{equation} which can be read from the general formula \eqref{ct} at $n=1$.
Therefore
\begin{equation}
\frac{1}{1-n}\log\frac{G(n)}{G(1)^n}=-\frac{1}{1-n}(nh_1-h_n)\frac{\pi^2}{6L^2}(x_{12})^2+O(x_{12}^3)\label{gng1}
\end{equation}
Plugging \eqref{gng1} into the expansion (\ref{Sncylinder2}), it is easy to check it indeed matches the direct calculation \eqref{Sncylinder} up to $O(x_{12}^3)$, when $\bar L=0$.  To obtain the result  \eqref{Sncylinder}  on an arbitrary spatial  circle, we need to perform one further spectral flow transformation with $p={\bar L\over L}$, and the final result is to simply replace $y$ with $y_{12}-px_{12}=y-{\bar L\over L}x_{12}$ as expected from  \eqref{Sncylinder}.

\subsection{The small $x$ expansion of  R\'enyi mutual information}

The normalization coefficients of the relevant quasi-primary operators can be found in the appendix \ref{OPEcoeff}. After using the summation formulae in the appendix \ref{Summation}, we find the R\'enyi mutual information of two disjoint intervals given by \eqref{Indef} and \eqref{Fn} can be expanded up to order $x^3$
\begin{eqnarray}
 \nonumber I_n&=& (\frac{2 b^2 (n+1)}{k n}+2 b\mu)x\\
 \nonumber  &&+ (\frac{b^2 (n+1)}{k n}+b\mu+\frac{c (n-1) (n+1)^2}{288
   n^3}+\frac{(n+1) \left(n^2+11\right)}{1440 n^3})x^2 \\
 \nonumber  && +(\frac{2 b^2 (n+1)}{3 k n}+\frac{2 b\mu}{3}+\frac{c (n-1)
   (n+1)^2}{288 n^3}+\frac{(n+1) \left(23 n^4+233
   n^2-40\right)}{30240 n^5})x^3\\
   && +O(x^4)
\end{eqnarray}
where we have kept the spectral flow parameter $\m$ arbitrary, but used the relation between the vacuum charges between $b$ and $a$.
Further using the vanishing charge condition at $n=1$, $b=-{k\mu\over2}$, we find that the RMI only depends on the central charge $c$ and the combination $a=-{b^2\over k}$, which will be specified by the theory itself.
Using the above relations, we can always rewrite the RMI as follows,
\be I_n=I_n^{(LO)}+I_n^{(NLO)}+O({1\over c})\label{mainresult}\ee
where the leading order result depends on both the central charge and the vacuum charge
\begin{eqnarray}
\nonumber  I_n^{(LO)}&=&  {2b^2 (n-1) \over k n}\big(x + {x^2\over 2} +{x^3\over 3}\big)\\
  && +\frac{c (n-1) (n+1)^2 }{288 n^3}(x^2+x^3) +O(x^4) \label{RMILO}
 \end{eqnarray}
whereas the next-to-leading-order result is of order $1$ and does not depend on  either the central charge or the vacuum charges.
 \begin{eqnarray}
 I_n^{(NLO)}&=& \frac{(n+1) \left(n^2+11\right)}{1440 n^3}x^2+\frac{(n+1) \left(23 n^4+233
   n^2-40\right)}{30240 n^5}x^3+O(x^4) \label{RMINLO}
\end{eqnarray}
The RMI for two disjoint intervals at large separation are hence given by \eqref{mainresult}-\eqref{RMINLO}, which are the main results of this paper.
We end this subsection with the following comments
\begin{itemize}
\item Up to level 3, there are no higher order terms at $o(1/c)$. This is reminiscent of  the observation in holographic CFTs {\cite{Chen:2013kpa}}  that two loop effects only can be seen from level four.  It will be interesting to find a simple explanation, or check it from a bulk calculation.
  \item The first line in (\ref{RMILO}) comes from the contribution of the operators $P_i$, while the remaining parts in \eqref{RMILO} are the same as that of the holomorphic part of a CFT$_2$.
  (\ref{RMILO}) suggests that the contributions from Virasoro and Kac-Moody sectors can factorize. This is a priori not obvious in
our explicit constructions of the quasi-primary operators level by level, as we were using a basis where the Virasoro and Kac-Moody sectors do not decouple. As will be shown in next subsection, factorization will be explicit using the method of conformal block expansion.
%Furthermore, in the computation of the one-point functions of the quasi-primary operators, we have introduced the warped conformal transformation (\ref{warpedtranf}), which is different from the usual uniformaization map in the literature. Nevertheless, the good agreement \note{between \eqref{RMILO} and (\ref{LO}) suggests that both are .}

  \item
  The reason for grouping the first line of (\ref{RMILO}) as the leading order result is that these terms come from the vacuum
  charges, instead of the Virasoro-Kac-Moody descendants. In the bulk we expect these terms to correspond to the on-shell action of a classical geometry.
  The vacuum charge $b$ depends on the theory, and will take different values in different examples of holography. For example, when the bulk dual is Einstein gravity on AdS$_3$ with the CSS boundary condition, there is a further relation between the parameters  \cite{Song:2016gtd} from $ L_0^{vac}\equiv a-{c\over24}=-{b^2\over k}-{c\over24}=0.$ Then it becomes obvious that (\ref{RMILO}) is indeed of order $c$ and corresponds to the classical contribution.
%\begin{equation}
%\m=\frac{1}{2}\sqrt{-\frac{l_{AdS}}{G_Nk}},\hs{2ex}b=\frac{1}{4}\sqrt{-\frac{l_{AdS}k}{G_N}},\hs{2ex}a=\frac{l_{AdS}}{16G_N},\hs{2ex}k=-\frac{l_{AdS}}{G_N},\label{parameters}
%\end{equation}
%\note{
 \item The next-to-leading order result \eqref{RMINLO}  up to $O(x^4)$ comes only from the Kac-Moody contribution. As will be seen later, this property can be reproduced in a bulk calculation.  This is different from the holomorphic part of holographic CFTs which is zero at this order \cite{Chen:2013kpa}.
  \end{itemize}

\section{R\'enyi mutual information from conformal blocks}
In this section we provide another way to read the R\'enyi mutual information at classical level by computing the four-point functions of the twist operators directly. In the orbifold WCFT, the central charge and Kac level is $nc$ and $nk$.
The striped four-point function of twist operators $F_n(x)$ can be written as a sum over the warped conformal blocks, and each block further factorizes as \cite{Fitzpatrick:2015zha,Apolo:2018eky} Kac-Moody block and Virasoro block,
\bea
F_n(x)&=& \frac{1}{x^{2h_n}}\sum_{p} C_{\sigma\tilde{\sigma} p}^2V_{T}(nc-1,h_n^{inv},h_p;x)V_{P}(nk,Q_n,x)
%&=&V_{T+P}(nc,h_n,nk,Q_n,h_p=0,q_p=0,x)+O(1)+O(\frac{1}{c})+\cdots
\eea
where $C_{\sigma\tilde{\sigma} p}$ is the OPE coefficient and the twist operator has the spectral flow invariant conformal weights $h_n^{inv}$ and U(1) charge given in \eqref{dinv}. The sum is over all primary operators labelled by $p$ with non-vanishing three point functions with $\sigma\tilde{\sigma}$.  By charge conservation, we learn that $q_p=0$.
%, and the spectral flow invariant weights $h_n^{inv}=h_n-{Q_n^2\over nk}$ and $h^{inv}_p\equiv h_p-\frac{q_p^2}{nk}$.
The Kac-Moody block is independent of the summation, and
 % and and $V_P$ is the Kac-Moody block.  $h_n^{inv}=h_n-\frac{Q_n^2}{nk}$.
%\begin{equation}
%V_{T+P}(c,h_i,k,q_i,h_p,q_p,x)=V_{T}(c-1,h_i-\frac{q_i^2}{k},h_p-\frac{q_p^2}{k})V_{P}(k,q_i,x),
%\end{equation}
has a closed form \cite{Fitzpatrick:2015zha},
\begin{equation}
V_{P}(k,q_i,x)=x^{\frac{2q_1^2}{k}}(1-x)^{\frac{-2q_1q_2}{k}}.
\end{equation}
 $F_n(x)$ can be written as a product of two parts
\begin{equation}
F_n(x)=(1-x)^{\frac{-2Q_n^2}{nk}} \{\frac{1}{x^{2h_n^{inv}}}\sum_{p} C_{\sigma\tilde{\sigma} p}^2V_{T}(nc-1,h_n^{inv},h_p;x)\}\label{fnf}
\end{equation}
The part in the braces is the same as the holomorphic part in a CFT \cite{Hartman:2013mia} with the shift of the central charge and conformal weights, while the remaining part  comes entirely from the Kac-Moody blocks.
For holographic WCFTs with a large $c$ and assuming that the vacuum block dominates the sum in \eqref{fnf},  the leading order contribution to the RMI \eqref{Indef} can be written as
\bea
 I^{(LO)}_n(x) &=&  \frac{2 b^2(n-1)}{k n}\log(1-x) \nn\\
 %&&+\frac{1}{2}(\mbox{R\'enyi mutual information of usual holographic CFT}). \\
&&% +\underbrace{{c(n+1)\over12 n}\log x+{1\over n-1}\log V_T(nc-1, \,h_n^{inv},0;x)}_{
+(\mbox{holomorphic part of RMI in holographic CFTs}).
  \label{LOpic}
\eea
where the second line comes from the  Virasoro block for the identity operator, and is exactly the same as the holomorphic part of RMI in holographic CFTs.

Now we check that \eqref{LOpic} agrees with our previous result \eqref{RMILO}, which requires the an explicit expression of the Virasoro block, which is in general not know in closed form. However, when the theory has a large $c$ expansion, the Virasoro block exponentiates as,
\begin{equation}
V_T(c,h_i,h_p;x)\sim\exp\{-\frac{c}{6}f(h_i,h_p,x)+O(1)+O(\frac{1}{c})+\cdots\}.
\end{equation}

Let us first consider the mutual information with $n=1$, $F_1(x)$ as defined in \eqref{fndef} can be read from a heavy-heavy-light-light (HHLL) four point function, which is dominated from vacuum block specified by $f_0(h_i,x)\equiv f(h_i,0,x)$, since the twist operators are light operators as $n\rightarrow 1$.
$f_0(h_i,x)$ can be calculated by the monodromy method in the Liouville field theory. The expression of $f_0(h_i,x)$ is an expansion in terms of the weights of light operators, %${h_L\over c}$,
but in closed form in terms of the weights of heavy operators and the cross ratio $x$\cite{Hartman:2013mia}. When the weights of the external operators are the same, the expansion of $f_0(h,x)$ reads\cite{Fitzpatrick:2014vua},
\begin{equation}
f_0(h,x)=\frac{12h}{c}\log x+o(\frac{h}{c}).
\end{equation}
This gives the mutual information $I_1$ when the two intervals are far away from each other,
\begin{equation}
I_1(x)=\frac{2 Q_1^2}{k}\log(1-x)=0,\hs{3ex}\mbox{if}\ x<\frac{1}{2}
\end{equation}
where $Q_1=0$ is the charge of the twist operator at $n=1$. This obviously agrees with \eqref{RMILO} for $n=1$, and is also consistent with the intuition that the mutual information for two disjoint intervals should vanish before reaching the point of phase transition.

More generally, for the  R\'enyi mutual information, it amounts to calculating the correlation function of four heavy operators, since the weights of twist operators grow with $c$ when $n\neq 1$. The Virasoro block at large $c$ can be expanded in terms of cross ratio $x$ by the recursion relations, no matter what the weights are\cite{Headrick:2010zt,Hartman:2013mia}
\begin{equation}
f_0(h,x)=12\delta \log x-3072\delta^2 q(x)^2+\frac{24576}{5}\delta^2(1-64\delta+704\delta^2)q(x)^4+\cdots
\end{equation}
where
\begin{equation}
\delta=\frac{h}{c},\ \ \ q(x)=e^{-\pi K(1-x)/K(x)}=\frac{1}{16}x+\frac{1}{32}x^2+\cdots
\end{equation}
and $K(x)$ is the complete elliptic integral of the first kind.
When the two intervals are far away from each other, the R\'enyi mutual information can be expanded in terms of $x$,
\begin{eqnarray}\label{LO}
\nonumber  I^{(LO)}_n(x) &=& \frac{2 b^2(n-1)}{k n}\log(1-x) \\
\nonumber  &&+\frac{c (n-1) (n+1)^2
   }{288 n^3}(x^2+x^3) \\
   &&+ \frac{c (n-1) (n+1)^2
   \left(1309 n^4-2
   n^2-11\right) x^4}{414720
   n^7}+O(x^5)
\end{eqnarray}
Note that the first line comes from the Kac-Moody block and is valid to all orders of $x$, while the second and third lines requires vacuum dominance and is valid at large $c$, and is a power series of $x$.
\eqref{LO} matches with the leading $c$ result in (\ref{RMILO}) up to $x^3$, which is a consistent check of the two methods.

\section{Holographic calculations from the bulk}
In this section we consider the holographic calculation of the R\'{e}nyi mutual information on AdS with the CSS boundary conditions in Einstein gravity.

Let us first review the story in the context of  AdS$_3$/CFT.
The bulk calculation of R\'{e}nyi entropy is to evaluate the partition function on a geometry whose asymptotic boundary, in the sense of Brown-Henneuax, is a given Riemann surface $\Sigma_n$.
All the classical configurations under Brown-Henneuax boundary conditions are the quotient of the global AdS$_3$ by a discrete subgroup $\Gamma$ of $SL(2,R)_L\times SL(2,R)_R$.
For technical reasons, it is usually more convenient to first consider the Euclidean theory and perform analytic continuation \cite{Krasnov:2000zq}. It is conjectured that the dominant contributions to the partition function are handlebody solutions which can be constructed using Schottky uniformation \cite{Faulkner:2013yia}.
After proper regularizations,  the on-shell action described by the Zograf-Takhtadzhyan(ZT) action\cite{ZTaction,Faulkner:2013yia}    agrees with the large $c$ result in the field theory calculation \cite{Headrick:2010zt,Hartman:2013mia,Chen:2013kpa}.
The 1-loop determinant \cite{Barrella:2013wja} agrees with the $O(c)$ result \cite{Headrick:2010zt,Chen:2013kpa}.

For AdS/WCFT, we expect a similar story, namely we expect the large $c$ result in the field theory calculation to match certain semiclassical calculation in the bulk, while the $O(c)$ is to match some 1-loop calculation.
We would borrow heavily techniques from AdS/CFT which are commonly formulated in Euclidean signature. On the other hand, we need to switch to Lorentzian signature temporarily as it will be easy to see the difference between CFT and WCFT. In section 3.1, we show how to  relate the Rinlder transformation \eqref{rindler} to the uniformization map \eqref{unip}. We proceed with the assumption that similar procedure can be taken in more general  cases with appropriate analytic continuation so that our results finds an explanation in Lorentzian signature.
In the following we first comment on the leading order contribution \eqref{RMILO} and then match the next-to-leading order result \eqref{RMINLO} from a the one-loop partition function in the bulk.
\subsection{Comments on the classical part}

Holographically, the large $c$ result in the WCFT calculation is expected to match certain semiclassical calculation in the bulk dual.
For single intervals, a holographic derivation of the entanglement entropy and R\'{e}nyi entropy has been performed in \cite{Song:2016gtd} in the context of WAdS/WCFT and AdS/WCFT. The bulk extension of the Rindler transformation \eqref{rindler} in Lorentzian signature can be explicitly constructed from a quotient  by elements of $SL(2, R)\times U(1)$. The novel feature for holographic entanglement entropy in the context of AdS/WCFT is that the spacelike geodesic in the bulk is attached to the boundary end points through null geodesics, which are along the bulk modular flow.

Similarly, we expect the above leading order result \eqref{LO} for RMI to correspond to the semiclassical action of the gravitational configuration. In Lorentzian signature, one may try to construct the bulk configurations whose asymptotic boundary is $\Sigma_n$ in the sense of CSS boundary conditions, and evaluate the on-shell action taking careful account of regularization. Schematically, the bulk construction of the $n$-smooth cover can be constructed  from quotients of  of $SL(2, R)\times U(1)$ and the Chiral Liouville action \cite{Compere:2013aya} might be useful to evaluate the on-shell action. One observation is that the contribution from the Virasoro block, i.e. the second and third line of \eqref{LO}, indeed agrees with a half of the result in ordinary holographic CFTs \cite{Chen:2013kpa}, which indeed agrees with the bulk calculation of the ZT action using the monodromy method\cite{Faulkner:2013yia}.
While it would be interesting to explicitly perform the complete bulk computation, we will leave it for future work.

\subsection{One-loop partition function in gravity}\label{oneloopbk}

In this subsection, we focus on the next-to-leading order result (\ref{RMINLO}), and show that it can be reproduced by the 1-loop quantum correction to the gravitational configuration in AdS$_3$ gravity with CSS boundary conditions.

Let us first review the story in the context of  AdS$_3$/CFT. All the classical configurations under Brown-Henneuax boundary conditions are quotients of the global AdS$_3$ by a discrete subgroup $\Gamma$ of $SL(2,R)_L\times SL(2,R)_R$.
For handlebody solutions,  the subgroup $\Gamma$ is a Schottky group, a finitely generated free group, such that all nontrivial elements are loxodromic. The 1-loop partition functions of the handle-body solutions with genus $g$ are studied in \cite{Yin:2007gv,Giombi:2008vd} by using the heat-kernel and the method of images
\bea
\label{1-loop} Z^{(1)}_{g;BH}&=&\prod_{\g\in\mP}Z_{BH}^{(1)}(q_\gamma, {\bar q}_\gamma)
\\&=&\prod_{\g\in\mP}Z_{BH}^{(1)}(q_\gamma) {\bar Z}^{(1)}_{BH}({\bar q}_\gamma).
\eea
where $\mP$ is a set of representatives of primitive conjugacy classes of the Schottky group $\G$. $q_\g$ is defined by writing the two eigenvalues of $\g \in \G$ as $q_\g^{\pm 1/2}$ with $|q_\g|<1$, and $Z_{BH}^{(1)}(q,\bar q)$ is the one loop determinant on a fixed saddle point solution with genus $1$ such as the BTZ black hole.
In the second line of the above equation, we have used the fact that the one loop contribution under the Brown-Hennaux boundary conditions factorizes into holomorphic part and anti-holomorphic part $Z_{BH}^{(1)}(q, \bar{q})=Z_{BH}^{(1)}(q) {\bar Z}_{BH}^{(1)}(\bar q)$. In general,  $\quad  Z_{BH}^{(1)}(q)=\prod_{s}\prod_{m=s}^\inf \f{1}{(1-q^m)}$, where the product over $s$ is with respect to the spins of massless fluctuations. For pure AdS$_3$ gravity with Brown-Henneaux boundary conditions, there is only massless graviton and $s=2$.

For AdS$_3$ gravity with the CSS boundary conditions, a similar story is expected to hold, namely we still expect that the generic 1-loop result on a fixed saddle point can be obtained from the image of BTZ result under the Schottky group. Namely, we still expect the general result \eqref{1-loop} to hold.
However, two crucial differences are expected due to the boundary conditions:  i) the 1-loop determinant around on BTZ black hole $Z^{(1)}_{   CSS}(q,\bar{q})$ and ii) the action of the Schottky group.

The 1-loop determinant for the BTZ black hole $Z^{(1)}_{CSS}(q,\bar{q})$ was studied in \cite{Castro:2017mfj}  using the quasi-normal mode method,
\be
Z^{(1)}_{CSS}(q,\bar q)=\prod_{l'=1}^\infty \frac{1}{1-q^{l'}}\prod_{l=2}^\infty \frac{1}{1-q^{l}}. \label{zbtz}
\ee
where $q$ is determined by the BTZ temperatures.
There are two remarkable points on this determinant. Firstly the determinant is the product of two factors, one for spin two and the other for spin one. Secondly, the fact that only $q$ appears in the final result indicates that only the left-mover survives the CSS boundary conditions. Therefore we will drop the dependence on $\bar q$ henceforth.
In fact, \eqref{zbtz} is just the contribution to the Virasoro-Kac-Moody character from the descendants \cite{Compere:2013bya,Apolo:2018eky}, which is consistent with the fact that the asymptotic symmetry is now generated by a Virasoro algebra and a Kac-Moody algebra, and that the holographic dual should be a WCFT.

For more general gravitational configurations compatible with the CSS boundary conditions, we expect that the 1-loop partition functions can be built from the BTZ result similar to \eqref{1-loop},
\be
Z^{(1)}_{g;CSS}=\prod_{\g\in\mP'}Z^{(1)}_{CSS}(q_\gamma)=\prod_{\gamma\in {\cal P}}\prod_{l'=1}^\infty \frac{1}{1-q_\gamma^{l'}}\prod_{l=2}^\infty \frac{1}{1-q_\gamma^{l}}. \label{1loop}
\ee
where $\mP'$ is a set of representatives of primitive conjugacy classes of the modified Schottky group $\G'$ which now should be compatible with the CSS boundary conditions.
Using one loop partition function given by \eqref{1loop}, the quantum correction to the holographic R\'{e}nyi mutual information at one-loop level reads,
 \begin{eqnarray}
\nonumber   I_n|_{1-loop} &=&-S_n(A\cup B)|_{1-loop}=\frac{-1}{1-n}(\log Z_n|_{1-loop}-n\log Z_1|_{1-loop}) \\
  &=&\frac{1}{n-1}\sum_{\gamma\in\mP'}\{\sum_{l=1}^{\infty}\log(1-q_{\gamma}^l)+\sum_{l=2}^{\infty}\log(1-q_{\gamma}^l)\}.\label{1loopbk}
 \end{eqnarray}
where we have used the fact that the holographic R\'{e}nyi entropy of single interval receives no quantum correction.

Now the key question is to determine the Schottky group $\G'$ that is compatible with the boundary conditions.
First note that the phase space of Einstein gravity under the CSS boundary conditions have the same fixed right moving energy. In particular, the zero mode solutions are BTZ black holes with fixed right moving energy.
All other configurations can be constructed from such BTZ black holes by first uncompactifying the non-contractable circle, and then identify points by elements of $SL(2,R)_L\times U(1)_R$, instead of the full fledged $SL(2,R)_L\times SL(2,R)_R$. Therefore we expect differences from the CFT results in the right moving sector.
Nevertheless, the symmetry algebra, the single interval uniformization map \eqref{unip} and the bulk Rindler transformation \cite{Song:2016gtd} strongly suggest that coordinate transformations in the left moving sector remain the same. Now we provide a heuristic argument.
Schematically, to calculate the RMI \eqref{1loopbk} in the bulk,
we need to find a $n$-th smooth cover $M_n$ in the bulk which is asymptotic, in the sense of CSS boundary conditions, to the $n$-sheeted geometry $\Sigma_n$. $M_n$ is a quotient of AdS$_3$, and can be obtained from a reference geometry by a local coordinate transformation. The reference geometry can be taken as the BTZ black hole with zero left-moving temperature, and fixed right-moving temperature. For example, with the conventions of \cite{Song:2016gtd}, the reference geometry is $T_u=0$ and a fixed value of $T_v$. We assume that the dominant contribution to the partition function is still given by the handle-body solutions which can be obtained from modified version of Schottky uniformization  which is compatible with the boundary conditions.
Using the holographic dictionary, the coordinate transformation from the reference geometry to $M_n$ will reduce to a warped conformal transformation in the form $z=f(w),\quad {\hat y}=y+g(z)$, where $(z,y')$ are coordinates on the plane, and $(w,y)$ are coordinates on $\Sigma_n$. Note that subtleties might appear due to the state dependence of the bulk to boundary map.
Furthermore, the boundary warped conformal map is determined by local properties near the branching points, as well as the global properties.
Near each branching point $w_i$,  the transformation can be expanded as
$z=(w-w_i)^{1\over n}, \,\hat y=y+n\mu \log  z$. Namely, they agree with the uniformization  \eqref{unip} for a single interval near all branching points.
The global properties will be determined by some monodromy conditions. For the $w$ direction, it will be the same monodromy conditions as in CFT$_2$ \cite{Faulkner:2013yia,Barrella:2013wja}.
As all the above requirements in the Virasoro direction $z$ is exactly the same as the holomorphic coordinate in a CFT$_2$, the action of the Schottky group also acts the same way in the holomorphic sector.
Therefore we expect to use group elements in \eqref{1loop} whose action on the right moving sector is different, while on the left moving sector is the same as in the Brown-Henneaux boundary conditions.  Since the BTZ 1-loop result is purely left moving, we expect no difference from the action of the usual Schottky group on the holomorphic sector.

With the above argument, in the following we will apply directly the 1-loop results of the holomorphic sector in AdS/CFT to AdS/WCFT,  and show that \eqref{1loopbk} indeed agrees with our next-to-leading result (\ref{RMINLO}) up to order $x^3$. In particular we will use $q_\gamma$'s from the previous study for double interval in the context of AdS/CFT, see
\cite{Barrella:2013wja} by Barrella et. al. In the following we give a brief review on the computation.

The Schottky uniformization  in the usual AdS/CFT case can be determined by the monodromy method. The single-valued coordinate $z$ on the $n$-sheeted geometry in the case with two disjoint intervals is the ratio of two independent solutions of the following differential equation,
 \begin{equation}
 \phi''(w)+\frac{1}{2}\sum_{i=1}^{4}(\frac{\Delta}{(w-w_i)}+\frac{\gamma_i}{w-w_i})\phi(w)=0,\hs{3ex}z=\frac{\phi_1(w)}{\phi_2(w)}.
 \end{equation}
 The map $z$ is single-valued with $\Delta=\frac{1}{2}(1-\frac{1}{n^2})$, since it is actually the series solution in the powers of $(w-w_i)^{1/n}$ at each point $w_i$, which is consistent with our uniformization formula \eqref{unip} for the left moving coordinate.
 When the solutions go around a closed loop enclosing branch points, they experience potential monodromy. The trivial monodromy at infinity fixes three of the accessory parameters $\gamma_i$, while the remaining one can be fixed by the monodromy around the trivial circles defining the Schottky group. The Schottky generators are respectively
 \begin{equation}
 L_1=M_2M_1,\hs{3ex}L_k=M_2^{k-1}L_1M_2^{-k+1}=M_2^kT^{-1}M_2TM_2^{-k+1},\hs{3ex}k=2,\cdots,n-1
 \end{equation}
 where $M_1,M_2$ are the monodromies obtained by encircling the branch points delimiting the nontrivial cycle and $T$ is the transformation matrix of the two basis of the power series solutions at these two points. All possible primitive words are made up of the Schottky generators and their inverses, up to conjugation in the Schottky group. There are infinite primitive classes for the cases with genus greater than one. Fortunately we are looking for the result  in terms of the power series of the cross ratio $x$. Consequently there are finite primitive classes contributing to a fixed power of $x$ in the partition function. Each pair of $T^{-1}$ and $T$ in the words gives the leading contribution of order $x^{2l}$ in the single term $\log(1-q^l)$. So at the leading order in $x$ expansion, the contribution comes from the so-called one-consecutively-decreasing words (1-CDW) with only one pair of $T^{-1}$ and $T$,
 \begin{equation}
 \gamma_{k,m}=L_{k+m}L_{k+m-1}\cdots L_{m+1}= M_2^{k+m}T^{-1}M_2^k TM_2^m.
 \end{equation}
For any given word length $k$ with $k\in[1,n-1]$, there are $n-k$ independent 1-CDWs, and the eigenvalues of $\gamma_{k,m}$ are independent of $m$.
 One of the eigenvalues of the 1-CDW above is vanishing, while the other one is
 \begin{equation}
 q^{-1/2}_{\gamma_{k}}\equiv q^{-1/2}_{\gamma_{k,m}}= -\frac{4n^2\sin^2\frac{k\pi}{n}}{x}+2+2(n^2-1)\sin^2\frac{k\pi}{n}+O(x). \label{eigen}
 \end{equation}
Up to $x^3$,  contributions to the 1-loop correction to the holographic R\'{e}nyi mutual information only come from the afarmentioned 1-CDW $\gamma_{k,m},$
 \begin{eqnarray}\label{spin2}
\nonumber   I_n|_{1-loop} &=&\frac{1}{n-1}\sum_{k=1}^{n-1}\sum_{m=1}^{n-k}\{\sum_{l=1}^\infty\log(1-q^l_{\gamma_{k}})+ \sum_{l=2}^{\infty}\log(1-q^l_{\gamma_{k}})\} +\sum_{other \,\gamma}\cdots\\
  &=&-\frac{1}{n-1} \sum_{k=1}^{n-1}(n-k)q_{\gamma_{k}}+ O(x^4).
 \end{eqnarray}
 In the first line, $\cdots$ denotes contributions from all other loxodromic element $\gamma$.
 From the above computation, we can directly see that there is no contributions from spin 2 up to $x^3$.
Combining with (\ref{eigen}),
  \begin{eqnarray}
\nonumber   I_n|_{1-loop} &=&-\frac{n}{2(n-1)} \sum_{k=1}^{n-1}\{\frac{s^{-4}_k}{16n^4}x^2+\frac{n^2s^{-4}_k+s^{-6}_k-s^{-4}_k}{16n^6}x^3+O(x^4)\}\\
  &=&\frac{(n+1) \left(n^2+11\right)}{1440 n^3}x^2+\frac{(n+1) \left(23 n^4+233
   n^2-40\right)}{30240 n^5}x^3+O(x^4), \label{Inbk}
 \end{eqnarray}
 where in the first line $s_k=\sin\frac{k\pi}{n}$.  This is indeed (\ref{RMINLO}).

 There are a few remarkable points:
\begin{enumerate}
\item Due to the CSS boundary conditions, the graviton in the bulk is ``chiral'', so only $q_\gamma$'s appear in the relation (\ref{1loop}).
\item From the sum in \eqref{spin2}, we can see that the contribution up to $x^3$ originates from the massless vector field. The contribution from spin 2 field will appear starting from $x^4$. This property also agrees with the WCFT result in section 4.4.
\item
The perfect matching between the WCFT result (\ref{RMINLO}) and the bulk result \eqref{Inbk} is very suggestive. To obtain the result \eqref{Inbk}, the above calculations only used the action of the Schottky group on the left moving sector, and is insensitive to the right movers. It supports that the Schottky group indeed acts the same way on the left-moving sector despite that we are using the CSS boundary conditions.
\item As mentioned before, we expect modifications to quotients in the $y$ direction due to the CSS boundary conditions. This is to be expected from the uniformization map \eqref{unip}, as well as the Rindler transformation in the bulk \cite{Song:2016gtd}. It would be interesting to find the action in the $U(1)$ direction, and show explicitly that it does not contribute to the 1-loop result.
\end{enumerate}

Another way to reproduce the 1-loop determinant (\ref{1loop}) from field theory is along the lines suggested in \cite{Chen:2015uga}. The essential point is that in the large $c$ limit, the conformal field theory becomes effectively free. After redefining the generators of the Virasoro-Kac-Moody algebra, we have
\be
[ {\hat L}_n, {\hat L}_m ] =\d_{n+m,0}, \hs{3ex}
[ {\hat L}_n, {\hat P}_m ] =0, \hs{3ex}
[ {\hat P}_n, {\hat P}_m ] =\d_{n+m,0}.
\ee
Therefore, in the large $c$ limit,  the states are generated not only by the Virasoro generators,  but also by the Kac-Moody generators. This is very similar to the case of CFT with $\mc{W}_3$ symmetry, but now there is a vector symmetry. Another subtle point is that now the symmetry has only holomorphic sector such that the partition function (\ref{1loop})  depends only on $q_\g$.

\section{Conclusion and Discussion }

In this work, we studied the R\'enyi mutual information of two disjoint intervals in the holographic warped CFT. In the semiclassical limit, we are allowed to
focus on the vacuum module of the theory.
We applied the OPE of the twist operators to collect the contributions of the quasi-primary operators up to level 3. In the large $c$ limit and the large distance limit, the result can be organized in powers of $1/c$ and the cross ratio $x$. The LO result is linear in $c$ and is expected to correspond to the classical action of the gravitational configuration. We performed a consistency check for the LO result by using the large $c$ expansion of the warped conformal block and found agreement up to $x^3$. More interestingly, we found that the LO R\'enyi mutual information can be closely related to the one in the usual holographic CFT by the relation (\ref{LOpic}): besides the contribution from the Kac-Moody block, which is of a logarithmic term, the remaining part is just half of the one in holographic CFT. It would certainly be important to understand this relation from the bulk gravity point of view.

The NLO result, which is independent of $c$, should correspond to the 1-loop partition function of the gravitational configuration according to the holographic dictionary. For a handlebody configuration, its 1-loop partition function can be rewritten in the form of (\ref{1loop}). As argued in section \ref{oneloopbk}, assuming the CSS boundary conditions do not affect the action of the quotient in the Virasoro direction,  we calculated 1-loop corrections to the R\'enyi mutual information and found agreement with the WCFT NLO result up to order $x^3$. This not only validates our treatment, but also provides nontrivial support to the  AdS$_3$/WCFT correspondence at 1-loop.

In the bulk calculation for the 1-loop correction, the CSS boundary conditions suggest that the gravitational configurations
are quotients of AdS$_3$ by elements from $SL(2,R)\times U(1)$. We provided an argument that the quotient acts the same way in the $x$ direction as in the usual AdS/CFT story. This allows us to borrow the results from previous calculations in AdS/CFT.  The agreement with the field theory result is indeed compatible with this argument. However, it would be nice to construct the gravitational configurations directly from quotients of AdS$_3$ by the group $SL(2,R)\times U(1)$, which respect the CSS boundary conditions in an obvious way. Such a construction has been done in \cite{Song:2016gtd} in order to calculate the entanglement entropy for a single interval. This is feasible by applying the coordinate transformation between arbitrary solutions in the CSS phase space metric similar to  \cite{Sheikh-Jabbari:2016znt}, and carefully matching to the boundary uniformization.

\vspace*{.5cm}
\noindent {\large{\bf Acknowledgments}} \\
We are grateful to Luis Apolo and Jian-fei Xu for valuable discussions.
The work of BC and PH were in part supported by NSFC Grant No.~11275010, No.~11335012, No.~11325522 and No. 11735001.
WS was supported by the National Thousand-Young-Talents Program of China and NFSC Grant No. 11735001. We thank  Tsinghua Sanya International Mathematics Forum for hospitality during the workshop ``Black holes and holography''.

\vspace*{1cm}
%\section*{Appendix}
\appendix
%\subsection{Notation}
\renewcommand{\appendixname}{Appendix~\Alph{section}}

\section{Coefficients and Normalization}\label{OPEcoeff}

In this appendix, we show how to calculate the one-point functions on the $n$-sheeted geometry in the single interval case. First, the one-point functions on the plane with non-trivial vacuum can be got by using the Ward identity and the normal order product. Then, we determine the transformation law of the composite operators and read out one-point functions on the $n$-sheeted geometry.

As an example, we calculate the coefficient of $A=T+\frac{c}{k}(P^2)$. The one-point function of $T$ in the non-trivial vacuum is,
\begin{equation}
\bra T(z) \ket_V=\lim_{y\rightarrow \infty,x\rightarrow 0}\frac{\bra V(y) T(z) V(x)\ket}{\bra V(y)V(x)\ket}=\frac{a}{z^2}
\end{equation}
where $V$ is the operator related with the non-trivial vacuum. The transformation law of $T$ is,
\begin{equation}
T(w)=w'^{-2}(T(z)-\frac{c}{12}s(w,z)-g'(z)P(z)-\frac{k g'(z)^2}{4})
\end{equation}
where $s(w,z)$ is the Schwarzian derivative, and $g(z)=n\mu\log z$. So the one-point function of $T$ on the $n$-sheeted geometry is,
\begin{equation}
\bra T(w)\ket_{\Sigma_n}=\frac{l^2}{w^2(w-l)^2}(\frac{c(n^2-1)+24a}{24n^2}-\frac{b\m}{n}-\frac{k\mu^2}{4}).
\end{equation}
Now, let us consider the composite operator $(P^2)$. The one-point function of $(P^2)$ on the plane with non-trivial vacuum is,
\begin{equation}
\bra(P^2)(z_2) \ket_V= \frac{1}{2\pi i}\oint\frac{1}{z_1-z_2}\bra P(z_1)P(z_2)\ket_V dz_1
\end{equation}
and
\begin{equation}
\bra P(z_1)P(z_2)\ket_V=\frac{b^2}{z_1z_2}-\frac{k}{2z_{12}^2}
\end{equation}
so
\begin{equation}
\bra (P^2)(z)\ket_V=\frac{b^2}{z^2}.
\end{equation}
The transformation law of $(P^2)$ should be  calculated by applying the normal order product procedure,
\begin{eqnarray}
\nonumber  (P^2)(w) &=& \frac{1}{2\pi i}\oint\frac{1}{w_1-w}P(w_1)P(w)dw_1 \\
\nonumber   &=&  \frac{1}{2\pi i}\oint dz_1\frac{1}{w(z_1)-w(z)}w'(z)^{-1}(P(z_1)+\frac{kg'(z_1)}{2})(P(z)+\frac{kg'(z)}{2})\\
\nonumber   &=& \frac{1}{2\pi i}\oint dz_1\frac{1}{w(z_1)-w(z)}w'(z)^{-1}(\frac{k}{2(z_1-z)^2}+(P^2)\\
\nonumber   &&+\frac{k}{2}P(z_1)g'(z)+\frac{k}{2}P(z)g'(z_1)+\frac{k^2}{4}g'(z_1)g'(z))\\
\nonumber   &=& \frac{k \left(3 w''(z)^2-2
   w^{(3)}(z) w'(z)\right)}{24
   w'(z)^4}+w'(z)^{-2}(P^2)(z)\\
   &&+\frac{b k g'(z)}{z w'(z)^2}+\frac{k^2 g'(z)^2}{4 w'(z)^2}
\end{eqnarray}
Using the explicit expression of the warped conformal transformation (\ref{unip}), we get
\begin{equation}
\bra P^2(w)\ket_{\Sigma_n}=\frac{l^2}{w^2(w-l)^2}(\frac{24b^2+k(1-n^2)}{24n^2}+\frac{bk\mu}{n}+\frac{k^2\m^2}{4}).
\end{equation}
Combining the results above, we find
\begin{equation}
\bra A(w)\ket_{\Sigma_n}=\frac{l^2}{w^2(w-l)^2}(\frac{a k+b^2 c}{k n^2}+\frac{b
   (c-1) \m}{n}+\frac{1}{4} (c-1) k
   \m^2).
\end{equation}
After taking the limit, the coefficient reads,
\begin{equation}
c_A=\frac{2 \left(\frac{a k+b^2 c}{k n^2}+\frac{b (c-1)
   \m}{n}+\frac{1}{4} (c-1) k \m^2\right)}{(c-1) c}=\frac{(2 b+k n\mu)^2}{2 c k n^2}.
\end{equation}

The above operator is defined in the same replica. For the composite operators with the operators on different replicas, their one-point functions are more tricky to compute. This is because that  the warped transformation (\ref{unip}) is not single-valued, nor does its derivatives. For example, consider the operator $P_iP_j$ with $i\neq j$. It transforms as
\begin{equation}
P_iP_j(w)=w_i'^{-1}w_j'^{-1}(P(z_i)+\frac{kg'(z_i)}{2})(P(z_j)+\frac{kg'(z_j)}{2})
\end{equation}
where $w_j$ is not single-valued. As $w\rightarrow \infty$,
\begin{equation}
w'_j(x)\rightarrow \frac{l}{n w^2}e^{2\pi i j/n}.
\end{equation}
Combining with the two-point function of $P(z_1)P(z_2)$ in the non-trivial vacuum, we get
\begin{equation}
c_{P_iP_j}=(\frac{4b^2}{k^2n^2}+\frac{1}{2kn^2s_{ij}^2}+\frac{4b\m}{kn}+\m^2).
\end{equation}
The other operators involved in our study can  be treated  in the similar manner.

The normalization factors of the quasi-primary operators can be computed straightforwardly by using the  algebra. For the quasi-primary operators appearing in our study, their normalization factors are listed as following
\begin{equation}
d_{-iP}=\frac{k}{2},\hs{3ex}d_T=\frac{c}{2},\hs{3ex}d_{P_iP_j}=\frac{k^2}{4},\hs{3ex}d_A=\frac{c(c-1)}{2},\end{equation}
\begin{equation}
d_{-iT_iP_j}=\frac{ck}{4},\hs{3ex}d_{iP_iP_jP_k}=\frac{k^3}{8},\hs{3ex}
d_{-iP_iA_j}=\frac{kc(c-1)}{4},
\end{equation}
\be
d_{P^3}=\frac{3k^3}{4},\hs{3ex}d_B=\frac{1}{4}k^3(c-1),\hs{3ex}d_{-iP_i\partial P_j+P_j\partial P_i}=k^2.
\ee

\section{Summation}\label{Summation}

In our computation, we  run into the following type of summations
\begin{equation}
\sum_{0\leq i\leq j<n}s_{ij}^{-m}=\frac{n}{2}f_m(n)=\frac{n}{2}\sum_{j=1}^{n-1}\frac{1}{(\sin \frac{j\pi }{n})^m}.
\end{equation}
Actually the function $f_m(n)$ can be calculated by using the inverse Mellin transformation,
\begin{equation}
\frac{1}{(\sin a\pi)^m}=\int_{0}^{\infty} x^{a-1}g_{\Delta}(x) dx
\end{equation}
where
\begin{equation}
g_{\Delta}(x)=\frac{2^{\Delta-2}}{\pi^2\Gamma[\Delta]\sqrt{x}}\Gamma[\frac{\Delta}{2}+i\frac{\log x}{2\pi}]\Gamma[\frac{\Delta}{2}-i\frac{\log x}{2\pi}].
\end{equation}
We can do the summation first and then calculate the integral to get the desired results.
There are two other kinds of summations involved in the calculation. They are respectively
\begin{equation}
\sum_{0\leq i<j<k\leq n-1}s_{ij}^{-2}+s_{ik}^{-2}+s_{jk}^{-2}=\frac{1}{6} (n-2) (n-1) n (n+1),
\end{equation}
\begin{equation}
\sum_{0\leq i<j<k\leq n-1}(s_{ij}^{-2}+s_{ik}^{-2}+s_{jk}^{-2})^2=\frac{1}{90} (n-2) (n-1) n (n+1)
   \left(n^2+8 n+27\right).
\end{equation}


\vspace*{5mm}
\begin{thebibliography}{99}


\bibitem{Maldacena:1997re}
J.~M. Maldacena, ``{The Large N limit of superconformal field theories and
  supergravity},''  Adv.Theor.Math.Phys. {\bfseries 2} (1998) 231--252,
\href{http://arxiv.org/abs/hep-th/9711200}{{\ttfamily arXiv:hep-th/9711200
  [hep-th]}}.\\
S.~Gubser, I.~R. Klebanov, and A.~M. Polyakov, ``{Gauge theory correlators from
  noncritical string theory},''
  \href{http://dx.doi.org/10.1016/S0370-2693(98)00377-3}{ Phys.Lett.
  {\bfseries B428} (1998) 105--114},
\href{http://arxiv.org/abs/hep-th/9802109}{{\ttfamily arXiv:hep-th/9802109
  [hep-th]}}.\\
E.~Witten, ``{Anti-de Sitter space and holography},''
  Adv.Theor.Math.Phys.{\bfseries 2} (1998) 253--291,
\href{http://arxiv.org/abs/hep-th/9802150}{{\ttfamily arXiv:hep-th/9802150
  [hep-th]}}.

  \bibitem{Ryu:2006bv}
  S.~Ryu and T.~Takayanagi,
  ``Holographic derivation of entanglement entropy from AdS/CFT,''
  Phys.\ Rev.\ Lett.\  {\bf 96}, 181602 (2006),
    [hep-th/0603001].

\bibitem{Ryu:2006ef}
  S.~Ryu and T.~Takayanagi,
  ``Aspects of Holographic Entanglement Entropy,''
  JHEP {\bf 0608}, 045 (2006)
    [hep-th/0605073].


\bibitem{Lewkowycz:2013nqa}
A.~Lewkowycz and J.~Maldacena, ``{Generalized gravitational entropy},''
  \href{http://dx.doi.org/10.1007/JHEP08(2013)090}{ JHEP {\bfseries 1308}
  (2013) 090},
\href{http://arxiv.org/abs/1304.4926}{{\ttfamily arXiv:1304.4926 [hep-th]}}.

  \bibitem{Faulkner:2013ana}
  T.~Faulkner, A.~Lewkowycz and J.~Maldacena,
  ``Quantum corrections to holographic entanglement entropy,''
  JHEP {\bf 1311}, 074 (2013)
   [arXiv:1307.2892 [hep-th]].

     \bibitem{Barrella:2013wja}
  T.~Barrella, X.~Dong, S.~A.~Hartnoll and V.~L.~Martin,
  ``Holographic entanglement beyond classical gravity,''
  JHEP {\bf 1309}, 109 (2013)
   [arXiv:1306.4682 [hep-th]].

\bibitem{Brown:1986nw}
J.~D. Brown and M.~Henneaux, ``{Central charges in the canonical realization of
  asymptotic symmetries: an example from three-dimensional gravity},''
\href{http://dx.doi.org/10.1007/BF01211590}{ Commun. Math. Phys.
  {\bfseries 104} (1986) 207--226}.

   \bibitem{Strominger:1997eq}
A.~Strominger, ``{Black hole entropy from near horizon microstates},''
  \href{http://dx.doi.org/10.1088/1126-6708/1998/02/009}{ JHEP {\bfseries
  9802} (1998) 009},
\href{http://arxiv.org/abs/hep-th/9712251}{{\ttfamily arXiv:hep-th/9712251
  [hep-th]}}.

   \bibitem{Hartman:2014oaa}
T.~Hartman, C.~A. Keller, and B.~Stoica, ``{Universal Spectrum of 2d Conformal
  Field Theory in the Large c Limit},''
  \href{http://dx.doi.org/10.1007/JHEP09(2014)118}{ JHEP {\bfseries 1409}
  (2014) 118},
\href{http://arxiv.org/abs/1405.5137}{{\ttfamily arXiv:1405.5137 [hep-th]}}.

\bibitem{Hartman:2013mia}
T.~Hartman, ``{Entanglement Entropy at Large Central Charge},''
\href{http://arxiv.org/abs/1303.6955}{{\ttfamily arXiv:1303.6955 [hep-th]}}.
%%CITATION = ARXIV:1303.6955;%%.

\bibitem{Faulkner:2013yia}
T.~Faulkner, ``{The Entanglement R\'enyi Entropies of Disjoint Intervals in
  AdS/CFT},''
\href{http://arxiv.org/abs/1303.7221}{{\ttfamily arXiv:1303.7221 [hep-th]}}.
%%CITATION = ARXIV:1303.7221;%%.


\bibitem{Headrick:2007km}
M.~Headrick and T.~Takayanagi, ``{A Holographic proof of the strong
  subadditivity of entanglement entropy},''
  \href{http://dx.doi.org/10.1103/PhysRevD.76.106013}{Phys. Rev.
  {\bfseries D76} (2007) 106013},
\href{http://arxiv.org/abs/0704.3719}{{\ttfamily arXiv:0704.3719 [hep-th]}}.

\bibitem{Hubeny:2007xt}
  V.~E.~Hubeny, M.~Rangamani and T.~Takayanagi,
  ``A Covariant holographic entanglement entropy proposal,''
  JHEP {\bf 0707}, 062 (2007)
 % %doi:10.1088/1126-6708/2007/07/062
  [arXiv:0705.0016 [hep-th]].

  \bibitem{Rangamani:2016dms}
  M.~Rangamani and T.~Takayanagi,
  ``Holographic Entanglement Entropy,''
  Lect.\ Notes Phys.\  {\bf 931}, pp.1 (2017)
  %doi:10.1007/978-3-319-52573-0
  [arXiv:1609.01287 [hep-th]].

\bibitem{Dong:2016fnf}
  X.~Dong,
  ``The Gravity Dual of R\'{e}nyi Entropy,''
  Nature Commun.\  {\bf 7}, 12472 (2016)
  %doi:10.1038/ncomms12472
  [arXiv:1601.06788 [hep-th]].

  \bibitem{Headrick:2010zt}
  M.~Headrick,
  ``Entanglement R\'{e}nyi entropies in holographic theories,''
  Phys.\ Rev.\ D {\bf 82}, 126010 (2010)
   [arXiv:1006.0047 [hep-th]].



 \bibitem{Chen:2013kpa}
  B.~Chen and J.~J.~Zhang,
  ``On short interval expansion of R\'enyi entropy,''
  JHEP {\bf 1311}, 164 (2013)
    [arXiv:1309.5453 [hep-th]].\\
B.~Chen, J.~Long and J.~j.~Zhang,
  ``Holographic R\'enyi entropy for CFT with W symmetry,''
  JHEP {\bf 1404}, 041 (2014)
    [arXiv:1312.5510 [hep-th]].

\bibitem{Headrick:2015gba}
  M.~Headrick, A.~Maloney, E.~Perlmutter and I.~G.~Zadeh,
  ``R\'enyi entropies, the analytic bootstrap, and 3D quantum gravity at higher genus,''
  JHEP {\bf 1507}, 059 (2015)
  %doi:10.1007/JHEP07(2015)059
  [arXiv:1503.07111 [hep-th]].

     \bibitem{OPE}
  E.~Perlmutter,
  ``Comments on R\'{e}nyi entropy in AdS$_3$/CFT$_2$,''
  JHEP {\bf 1405}, 052 (2014)
    [arXiv:1312.5740 [hep-th]].
 B.~Chen, F.~y.~Song and J.~j.~Zhang,
  ``Holographic R\'enyi entropy in AdS$_3$/LCFT$_2$ correspondence,''
  JHEP {\bf 1403}, 137 (2014)
    [arXiv:1401.0261 [hep-th]].
    M.~Beccaria and G.~Macorini,
  ``On the next-to-leading holographic entanglement entropy in $AdS_{3}/CFT_{2}$,''
  JHEP {\bf 1404}, 045 (2014)
   [arXiv:1402.0659 [hep-th]].
  J.~j.~Zhang,
  ``Holographic R\'enyi entropy for two-dimensional $ \mathcal{N}=\left(1,\;1\right) $ superconformal field theory,''
  JHEP {\bf 1512}, 027 (2015)
   [arXiv:1510.01423 [hep-th]].
  Z.~Li and J.~j.~Zhang,
  ``On one-loop entanglement entropy of two short intervals from OPE of twist operators,''
  JHEP {\bf 1605}, 130 (2016)
    [arXiv:1604.02779 [hep-th]].
    Z.~Li and J.~j.~Zhang,
  ``Holographic R\'enyi entropy for two-dimensional $\mathcal{N}$=(2,2) superconformal field theory,''
  arXiv:1611.00546 [hep-th].

  \bibitem{Chen:2014unl}	B.~Chen and J.~q.~Wu,
		``Single interval R\'enyi entropy at low temperature,''
		JHEP {\bf 1408}, 032 (2014)
		[arXiv:1405.6254 [hep-th]].\\
    B.~Chen and J.~q.~Wu,
  ``Holographic calculation for large interval R\'enyi entropy at high temperature,''
  Phys.\ Rev.\ D {\bf 92}, no. 10, 106001 (2015)
    [arXiv:1506.03206 [hep-th]].\\
    B.~Chen, J.~B.~Wu and J.~j.~Zhang,
  ``Short interval expansion of R\'enyi entropy on torus,''
  JHEP {\bf 1608}, 130 (2016)
    [arXiv:1606.05444 [hep-th]].

 \bibitem{Belin:2017nze}
  A.~Belin, C.~A.~Keller and I.~G.~Zadeh,
  ``Genus two partition functions and Rényi entropies of large c conformal field theories,''
  J.\ Phys.\ A {\bf 50}, no. 43, 435401 (2017)
  doi:10.1088/1751-8121/aa8a11
  [arXiv:1704.08250 [hep-th]].
  %%CITATION = doi:10.1088/1751-8121/aa8a11;%%
  %18 citations counted in INSPIRE as of 22 May 2019

%\cite{Dong:2018esp}
\bibitem{Dong:2018esp}
  X.~Dong, S.~Maguire, A.~Maloney and H.~Maxfield,
  ``Phase transitions in 3D gravity and fractal dimension,''
  JHEP {\bf 1805}, 080 (2018)
  doi:10.1007/JHEP05(2018)080
  [arXiv:1802.07275 [hep-th]].
  %%CITATION = doi:10.1007/JHEP05(2018)080;%%
  %5 citations counted in INSPIRE as of 22 May 2019

\bibitem{Casini:2011kv}
  H.~Casini, M.~Huerta and R.~C.~Myers,
  ``Towards a derivation of holographic entanglement entropy,''
  JHEP {\bf 1105}, 036 (2011)
  %%doi:10.1007/JHEP05(2011)036
  [arXiv:1102.0440 [hep-th]].
  %%CITATION = %doi:10.1007/JHEP05(2011)036;%%
  %645 citations counted in INSPIRE as of 20 Mar 2019



\bibitem{Compere:2013bya}
  G.~Comp\'{e}re, W.~Song and A.~Strominger,
  ``New Boundary Conditions for AdS3,''
  JHEP {\bf 1305}, 152 (2013)
  %doi:10.1007/JHEP05(2013)152
  [arXiv:1303.2662 [hep-th]].

\bibitem{Hofman:2011zj}
  D.~M.~Hofman and A.~Strominger,
  ``Chiral Scale and Conformal Invariance in 2D Quantum Field Theory,''
  Phys.\ Rev.\ Lett.\  {\bf 107}, 161601 (2011)
  %doi:10.1103/PhysRevLett.107.161601
  [arXiv:1107.2917 [hep-th]].

 \bibitem{Detournay:2012pc}
  S.~Detournay, T.~Hartman and D.~M.~Hofman,
  ``Warped Conformal Field Theory,''
  Phys.\ Rev.\ D {\bf 86}, 124018 (2012)
  %doi:10.1103/PhysRevD.86.124018
  [arXiv:1210.0539 [hep-th]].

\bibitem{Castro:2015csg}
  A.~Castro, D.~M.~Hofman and N.~Iqbal,
  ``Entanglement Entropy in Warped Conformal Field Theories,''
  JHEP {\bf 1602}, 033 (2016)
  %doi:10.1007/JHEP02(2016)033
  [arXiv:1511.00707 [hep-th]].

  \bibitem{Song:2016gtd}
  W.~Song, Q.~Wen and J.~Xu,
  ``Modifications to Holographic Entanglement Entropy in Warped CFT,''
  JHEP {\bf 1702}, 067 (2017)
  %doi:10.1007/JHEP02(2017)067
  [arXiv:1610.00727 [hep-th]].

  \bibitem{Song:2017czq}
  W.~Song and J.~Xu,
  ``Correlation Functions of Warped CFT,''
  JHEP {\bf 1804}, 067 (2018)
  %doi:10.1007/JHEP04(2018)067
  [arXiv:1706.07621 [hep-th]].

\bibitem{nielsen2010quantum}
M.~A. Nielsen and I.~L. Chuang, {\em Quantum computation and quantum
  information}.
\newblock Cambridge university press, 2010.

\bibitem{petz2008quantum}
D.~Petz, {\em Quantum information theory and quantum statistics}.
\newblock Springer, 2008.

%\cite{Jiang:2017ecm}
\bibitem{Jiang:2017ecm}
  H.~Jiang, W.~Song and Q.~Wen,
  ``Entanglement Entropy in Flat Holography,''
  JHEP {\bf 1707}, 142 (2017)
  %%doi:10.1007/JHEP07(2017)142
  [arXiv:1706.07552 [hep-th]].
  %%CITATION = %doi:10.1007/JHEP07(2017)142;%%
  %20 citations counted in INSPIRE as of 13 Mar 2019


%\cite{Azeyanagi:2018har}
\bibitem{Azeyanagi:2018har}
  T.~Azeyanagi, S.~Detournay and M.~Riegler,
  ``Warped Black Holes in Lower-Spin Gravity,''
  Phys.\ Rev.\ D {\bf 99}, no. 2, 026013 (2019)
  %%doi:10.1103/PhysRevD.99.026013
  [arXiv:1801.07263 [hep-th]].
  %%CITATION = %doi:10.1103/PhysRevD.99.026013;%%
  %5 citations counted in INSPIRE as of 15 Feb 2019



\bibitem{Holzhey:1994we}
  C.~Holzhey, F.~Larsen and F.~Wilczek,
  ``Geometric and renormalized entropy in conformal field theory,''
  Nucl.\ Phys.\ B {\bf 424}, 443 (1994)
   [hep-th/9403108].

  \bibitem{Calabrese:2004eu}
  P.~Calabrese and J.~L.~Cardy,
  ``Entanglement entropy and quantum field theory,''
  J.\ Stat.\ Mech.\  {\bf 0406}, P06002 (2004)
    [hep-th/0405152].

  \bibitem{Calabrese:2005zw}
  P.~Calabrese and J.~L.~Cardy,
  ``Entanglement entropy and quantum field theory: A Non-technical introduction,''
  Int.\ J.\ Quant.\ Inf.\  {\bf 4}, 429 (2006)
    [quant-ph/0505193].


\bibitem{Wolf}
M.M.~Wolf, F.~Verstraete, M.B.~Hastings, J.I.~Cirac, ``Area laws in quantum systems: mutual information and correlations", Phys.\ Rev.\ Lett.,{\bf 100},070502 (2008) [arXiv:0704.3906 [quant-ph]].



 \bibitem{Calabrese:2010he}
  P.~Calabrese, J.~Cardy and E.~Tonni,
  ``Entanglement entropy of two disjoint intervals in conformal field theory II,''
  J.\ Stat.\ Mech.\  {\bf 1101}, P01021 (2011)
   [arXiv:1011.5482 [hep-th]].



\bibitem{Apolo:2018oqv}
  L.~Apolo, S.~He, W.~Song, J.~Xu and J.~Zheng,
  ``Entanglement and chaos in warped conformal field theories,''
  arXiv:1812.10456 [hep-th].


\bibitem{Perlmutter:2016pkf}
  E.~Perlmutter,
  ``Bounding the Space of Holographic CFTs with Chaos,''
  JHEP {\bf 1610}, 069 (2016)
  doi:10.1007/JHEP10(2016)069
  [arXiv:1602.08272 [hep-th]].
  %%CITATION = doi:10.1007/JHEP10(2016)069;%%
  %79 citations counted in INSPIRE as of 22 May 2019


  \bibitem{Fitzpatrick:2015zha}
  A.~L.~Fitzpatrick, J.~Kaplan and M.~T.~Walters,
  ``Virasoro Conformal Blocks and Thermality from Classical Background Fields,''
  JHEP {\bf 1511}, 200 (2015)
  %doi:10.1007/JHEP11(2015)200
  [arXiv:1501.05315 [hep-th]].

 \bibitem{Apolo:2018eky}
  L.~Apolo and W.~Song,
  ``Bootstrapping holographic warped CFTs or: how I learned to stop worrying and tolerate negative norms,''
  JHEP {\bf 1807}, 112 (2018)
  %doi:10.1007/JHEP07(2018)112
  [arXiv:1804.10525 [hep-th]].

  \bibitem{Krasnov:2000zq}
  K.~Krasnov,
  ``Holography and Riemann surfaces,''
  Adv.\ Theor.\ Math.\ Phys.\  {\bf 4}, 929 (2000)
  %doi:10.4310/ATMP.2000.v4.n4.a5
  [hep-th/0005106].

\bibitem{ZTaction}
P. G.~Zograf and L.A.~Takhtadzhyan, ``On the uniformization of Riemann surfaces and the Weil-Petersson metric on TeichmÃŒller and Schottky spaces",  1988. Math.USSR Sb.,60,297.

 \bibitem{Compere:2013aya}
  G.~Comp\'{e}re, W.~Song and A.~Strominger,
  ``Chiral Liouville Gravity,''
  JHEP {\bf 1305}, 154 (2013)
  %%doi:10.1007/JHEP05(2013)154
  [arXiv:1303.2660 [hep-th]].

 \bibitem{Yin:2007gv}
  X.~Yin,
  ``Partition Functions of Three-Dimensional Pure Gravity,''
  Commun.\ Num.\ Theor.\ Phys.\  {\bf 2}, 285 (2008)
  [arXiv:0710.2129 [hep-th]].

\bibitem{Giombi:2008vd}
S.~Giombi, A.~Maloney, and X.~Yin, ``{One-loop Partition Functions of 3D
  Gravity},'' \href{http://dx.doi.org/10.1088/1126-6708/2008/08/007}{{ JHEP} {\bfseries 0808} (2008) 007},
\href{http://arxiv.org/abs/0804.1773}{{\ttfamily arXiv:0804.1773 [hep-th]}}.

  \bibitem{Castro:2017mfj}
  A.~Castro, C.~Keeler and P.~Szepietowski,
  ``Tweaking one-loop determinants in AdS$_{3}$,''
  JHEP {\bf 1710}, 070 (2017)
  %doi:10.1007/JHEP10(2017)070
  [arXiv:1707.06245 [hep-th]].




%\cite{Compere:2009zj}
\bibitem{Compere:2009zj}
  G.~Compere and S.~Detournay,
  ``Boundary conditions for spacelike and timelike warped AdS$_3$ spaces in topologically massive gravity,''
  JHEP {\bf 0908}, 092 (2009)
  %doi:10.1088/1126-6708/2009/08/092
  [arXiv:0906.1243 [hep-th]].
  %%CITATION = %doi:10.1088/1126-6708/2009/08/092;%%
  %82 citations counted in INSPIRE as of 19 Mar 2019





\bibitem{Chen:2015uga}
B.~Chen and J.-q. Wu, ``{1-loop partition function in AdS$_{3}$/CFT$_{2}$},''
  \href{http://dx.doi.org/10.1007/JHEP12(2015)109}{{ JHEP} {\bfseries 1512}
  (2015) 109},
\href{http://arxiv.org/abs/1509.02062}{{\ttfamily arXiv:1509.02062 [hep-th]}}.
%%CITATION = ARXIV:1509.02062;%%.

\bibitem{Sheikh-Jabbari:2016znt}
  M.~M.~Sheikh-Jabbari and H.~Yavartanoo,
  ``Excitation entanglement entropy in two dimensional conformal field theories,''
  Phys.\ Rev.\ D {\bf 94}, no. 12, 126006 (2016)
  %doi:10.1103/PhysRevD.94.126006
  [arXiv:1605.00341 [hep-th]].



\bibitem{Grumiller:2016pqb}
 C.~Troessaert,
  ``Enhanced asymptotic symmetry algebra of $AdS$$_{3}$,''
  JHEP {\bf 1308}, 044 (2013)
  %doi:10.1007/JHEP08(2013)044
  [arXiv:1303.3296 [hep-th]].
  \\  S.~G.~Avery, R.~R.~Poojary and N.~V.~Suryanarayana,
  ``An sl(2,$\mathbb{R}$) current algebra from $AdS_3$ gravity,''
  JHEP {\bf 1401}, 144 (2014)
  %doi:10.1007/JHEP01(2014)144
  [arXiv:1304.4252 [hep-th]].
  \\ L.~Apolo and M.~Porrati,
  ``Free boundary conditions and the AdS$_3$/CFT$_2$ correspondence,''
  JHEP {\bf 1403}, 116 (2014)
  %doi:10.1007/JHEP03(2014)116
  [arXiv:1401.1197 [hep-th]].
\\H.~Afshar, S.~Detournay, D.~Grumiller, W.~Merbis, A.~Perez, D.~Tempo and R.~Troncoso,
  ``Soft Heisenberg hair on black holes in three dimensions,''
  Phys.\ Rev.\ D {\bf 93}, no. 10, 101503 (2016)
  %doi:10.1103/PhysRevD.93.101503
  [arXiv:1603.04824 [hep-th]].
  \\  A.~P\'{e}rez, D.~Tempo and R.~Troncoso,
  ``Boundary conditions for General Relativity on AdS$_{3}$ and the KdV hierarchy,''
  JHEP {\bf 1606}, 103 (2016)
  %doi:10.1007/JHEP06(2016)103
  [arXiv:1605.04490 [hep-th]].
 \\ D.~Grumiller and M.~Riegler,
  ``Most general AdS$_{3}$ boundary conditions,''
  JHEP {\bf 1610}, 023 (2016)
  %doi:10.1007/JHEP10(2016)023
  [arXiv:1608.01308 [hep-th]].


  \bibitem{Asplund:2014coa}
  C.~T.~Asplund, A.~Bernamonti, F.~Galli and T.~Hartman,
  ``Holographic Entanglement Entropy from 2d CFT: Heavy States and Local Quenches,''
  JHEP {\bf 1502}, 171 (2015)
  %doi:10.1007/JHEP02(2015)171
  [arXiv:1410.1392 [hep-th]].

  \bibitem{Fitzpatrick:2014vua}
  A.~L.~Fitzpatrick, J.~Kaplan and M.~T.~Walters,
  ``Universality of Long-Distance AdS Physics from the CFT Bootstrap,''
  JHEP {\bf 1408}, 145 (2014)
  %doi:10.1007/JHEP08(2014)145
  [arXiv:1403.6829 [hep-th]].


\bibitem{yu}
B.~Yu, ``Spaceness Conditions For Warped CFTs, '' in progress


%\cite{Dong:2016hjy}
\bibitem{Dong:2016hjy}
  X.~Dong, A.~Lewkowycz and M.~Rangamani,
  ``Deriving covariant holographic entanglement,''
  JHEP {\bf 1611}, 028 (2016)
 % doi:10.1007/JHEP11(2016)028
  [arXiv:1607.07506 [hep-th]].
  %%CITATION = doi:10.1007/JHEP11(2016)028;%%
  %73 citations counted in INSPIRE as of 26 Mar 2019

%\cite{Wen:2018mev}
\bibitem{Wen:2018mev}
  Q.~Wen,
  ``Towards the generalized gravitational entropy for spacetimes with non-Lorentz invariant duals,''
  JHEP {\bf 1901}, 220 (2019)
  %doi:10.1007/JHEP01(2019)220
  [arXiv:1810.11756 [hep-th]].
  %%CITATION = doi:10.1007/JHEP01(2019)220;%%
  %4 citations counted in INSPIRE as of 26 Mar 2019

\end{thebibliography}
\end{document}